\documentclass{amsart}
\usepackage{hyperref}
\usepackage{graphicx}
\usepackage{curves}
\usepackage{float}
\usepackage{caption}
\usepackage{subcaption}
\usepackage{dirtytalk}
\usepackage{amsthm}
\usepackage{amsbsy}
\theoremstyle{plain}

\newcommand*{\mailto}[1]{\href{mailto:#1}{\nolinkurl{#1}}}

\newtheorem{theorem}{Theorem}[section]

\newtheorem{remark}[theorem]{Remark}

\newtheorem*{src*}{Soliton Resolution Conjecture}

\newcommand{\R}{{\mathbb R}}
\newcommand{\N}{{\mathbb N}}
\newcommand{\Z}{{\mathbb Z}}
\newcommand{\C}{{\mathbb C}}
\newcommand{\M}{{\mathbb M}}

\newcommand{\nn}{\nonumber}
\newcommand{\be}{\begin{equation}}
\newcommand{\ee}{\end{equation}}
\newcommand{\bea}{\begin{eqnarray}}
\newcommand{\eea}{\end{eqnarray}}
\newcommand{\ul}{\underline}

\newcommand{\ti}{\tilde}

\newcommand{\spr}[2]{\langle #1 , #2 \rangle}

\newcommand{\I}{\mathrm{i}}

\newcommand{\re}{\mathrm{Re}}

\def\XXint#1#2#3{{\setbox0=\hbox{$#1{#2#3}{\int}$}
      \vcenter{\hbox{$#2#3$}}\kern-.5\wd0}}

\newcommand{\ulz}{\underline{z}}
\newcommand{\di}{\mathcal{D}}
\newcommand{\vrc}{\ul{\Xi}_{p_0}}
\newcommand{\hvrc}{\ul{\hat{\Xi}}_{p_0}}
\newcommand{\hmu}{\hat{\mu}}

\newcommand{\dimuz}{\di_{\ul{\hat{\mu}}}}

\newcommand{\Amap}{\ul{A}_{p_0}}
\newcommand{\amap}{\ul{\alpha}_{p_0}}
\newcommand{\hAmap}{\ul{\hat{A}}_{p_0}}
\newcommand{\hamap}{\ul{\hat{\alpha}}_{p_0}}

\newcommand{\Rg}[1]{R_{2g+2}^{1/2}(#1)}

\newcommand{\eps}{\varepsilon}

\newcommand{\sig}{\sigma}
\newcommand{\lam}{\lambda}
\newcommand{\gam}{\gamma}
\newcommand{\om}{\omega}

\numberwithin{equation}{section}

\begin{document}

\title[On Soliton Resolution for a Lattice]{On Soliton Resolution  for a Lattice}

\author[N. Hatzizisis]{Nicholas Hatzizisis $^{\dag}$}
\address{$^{\dag}$ Mathematics Building, University of Crete \\
700 13 Voutes, Greece}
\email{\mailto{nhatzitz@gmail.com}}
\urladdr{\url{https://nikoshatzizisis.wordpress.com/home/}}

\author[S. Kamvissis]{Spyridon Kamvissis $^{\ddag}$}
\address{$^{\ddag}$ Mathematics Building, University of Crete \\
700 13 Voutes, Greece}
\email{\mailto{spyros@tem.uoc.gr}}
\urladdr{\url{http://www.tem.uoc.gr/~spyros/}}

\thanks{Research supported in part by grant 10331/2019
at the University of Crete.}

\keywords{Soliton Resolution, FPUT lattice}
\subjclass[2000]{Primary 37K40, 37K45; Secondary 35Q15, 37K10}

\bigskip

\begin{abstract}
The soliton resolution conjecture for evolution PDEs of dispersive type states (vaguely) that
generic initial data of finite energy
give rise asymptotically to a set of receding solitons and a decaying background radiation.

In this letter, we investigate a  possible extension of  this conjecture to  
discrete lattices of the Fermi-Pasta-Ulam-Tsingou type (rather than PDEs) in two cases; the case with
initial data of finite energy and a more general case with initial data 
that are a short range perturbation of a periodic function. 

In the second case, inspired by rigorous results on the Toda lattice, 
we suggest that the soliton resolution phenomenon is 
replaced by something somewhat more complicated: a short range perturbation of
a periodic function actually gives rise to different phenomena in different regions.
Apart from regions of (asymptotically) pure periodicity and regions of solitons in a periodic background, 
we also observe
“modulated” regions of fast oscillations with slowly varying parameters like amplitude and phase.

We have conducted some  numerical calculations to investigate if this trichotomy 
(pure periodicity + solitons + modulated) persists for any
discrete lattices of the Fermi-Pasta-Ulam-Tsingou type.  For small perturbations of integrable lattices like the
linear harmonic lattice, the Langmuir chain and the Toda lattice, this is true. But in general even
chaotic phenomena can occur.

\end{abstract}

\maketitle

\section{Historical Introduction and a Statement of the Soliton Resolution Conjecture in a Periodic Background}

A classical observation going back to the seminal discovery of
Zabusky and Kruskal (\cite{zakr}) states that a local (or \say{short range})  perturbation
of the trivial stationary solution of a completely integrable soliton PDE (like KdV or NLS) or lattice (like Toda),
eventually splits into a number of receeding solitons plus a (uniformly) decaying \say{background radiation}.
The first complete description for the long time asymptotics of the KdV equation were given in 
\cite{as}. Rigorous proofs  can be constructed for any system solvable via the inverse scattering theory.
Such proofs routinely employ  the asymptotic analysis of the associated Riemann-Hilbert factorisation problems,
at least in the case of one space dimension (\cite{dz}, \cite{dvz}, \cite{k1}, \cite{k2}), 
where the inverse scattering problem is equivalent
to a Riemann-Hilbert factorisation problem in the complex plane.

More recently, an even more daring conjecture has begun to take shape (\cite{t}, \cite{c}):
for $any$ dispersive PDE of NLS or KdV type
in any spatial dimension (!),
generic initial data of bounded energy
give rise asymptotically to a set of receeding solitons and a decaying background radiation.

Our aim here is
to investigate the Soliton Resolution Conjecture for one-dimensional, constant or periodic background, 
uniform (without impurities, i.e. all particles
are of the same mass $m=1$), doubly infinite lattices with nearest neighbor interaction (each particle only
affects its two neighbors, one on the left and one on the right).
 
To be precise, let $x_n(t)=x(n,t)$, $(n,t) \in \Z \times \R$ denote the displacement (from its equilibrium) 
of the $n^{th}$ particle in the chain at time $t$. If we denote by $V(x_{n+1}-x_n)$, $n\in\mathbb{Z}$ 
the interaction potential between neighboring particles, then the equation of motion is given by
\be \label{FPUT}
\aligned
\frac{d^2x_n}{dt^2} = V'(x_{n+1}-x_n)-V'(x_n-x_{n-1}),\quad (n,t) \in \Z \times \R
\endaligned
\ee
where $V'(x)=\frac{dV}{dx}(x)=:-F(x)$, $V$ being  the potential function and $F$ the
corresponding force. What can we say about the long time asymptotics of this system given some general conditions
on the behaviour of the initial data $x_n$, and $dx_n/dt$ at time $t=0$ and as $n \to \pm \infty$?

Let us begin by presenting  a Soliton Resolution Conjecture in a constant background. To be more precise,
 we assume that $x_n$, and $dx_n/dt$ tend to  $0$ fast enough as $n \to \pm \infty$ (at time $0$).
(See (A.3)n the Appendix A, for a definition of "fast enough" in the case of the Toda lattice,
where rigorous results exist. Even somewhat weaker definitons are probably sufficent.) 
The claim is that the solution is asymptotically given by a sum of solitary waves with different speeds plus
a small "radiation" term that decays in time.

The first rigorous study of this phenomenon in a constant background was done in 
\cite{k1} for the special case of
the Toda lattice, with $V(x)=e^{-x}+x$,
in the case where the associated Jacobi operatorm has no eigenvalues.
Eigenvalues were added later in \cite{krt2}.
In these works it was shown that the error term is actually of order $O(t^{-1/2})$ uniformly in $n$,
at least away from the two regions where $n/t $ is $\pm 1+ o(1)$. With some more work, one can  actually 
show that in these small regions the error order is  $O(t^{-1/3})$.

The first rigorous study of the analogous phenomenon in a periodic background was done in \cite{kt1}, 
also for the  Toda lattice,
where numerical experiments were presented and complete analytic formulas where given for the asymptotics of the 
doubly infinite periodic Toda lattice under a "short range" perturbation (again see appendices \ref{asympt} and 
\ref{secAG} for the exact condition on the initial data and the exact asymptotic formulae). The proofs were 
presented in \cite{kt2} in the case where the associated Lax operator 
(tridiagonal Jacobi operator in this case)
has no eigenvalues. Again, one uses asymptotic analysis of the associated Riemann-Hilbert holomorphic 
factorisation problems,
with the extra novelty that such problems are posed on a Riemann surface. 
Once this analysis was achieved, eigenvalues 
were easily added (\cite{krt})
\footnote{Eigenvalues turn the associated Riemann-Hilbert 
factorisation problems into meromorphic problems, but simple tricks (\cite{dkkz}) can change such 
problems back into holomorphic problems which can be asymptotically analysed after some transformations.} 
and higher order asymptotics have also been presented (\cite{kt2}).

\begin{figure}[H]
    \centering
    \begin{subfigure}[b]{0.48\linewidth}        
        \centering
        \includegraphics[width=\linewidth]{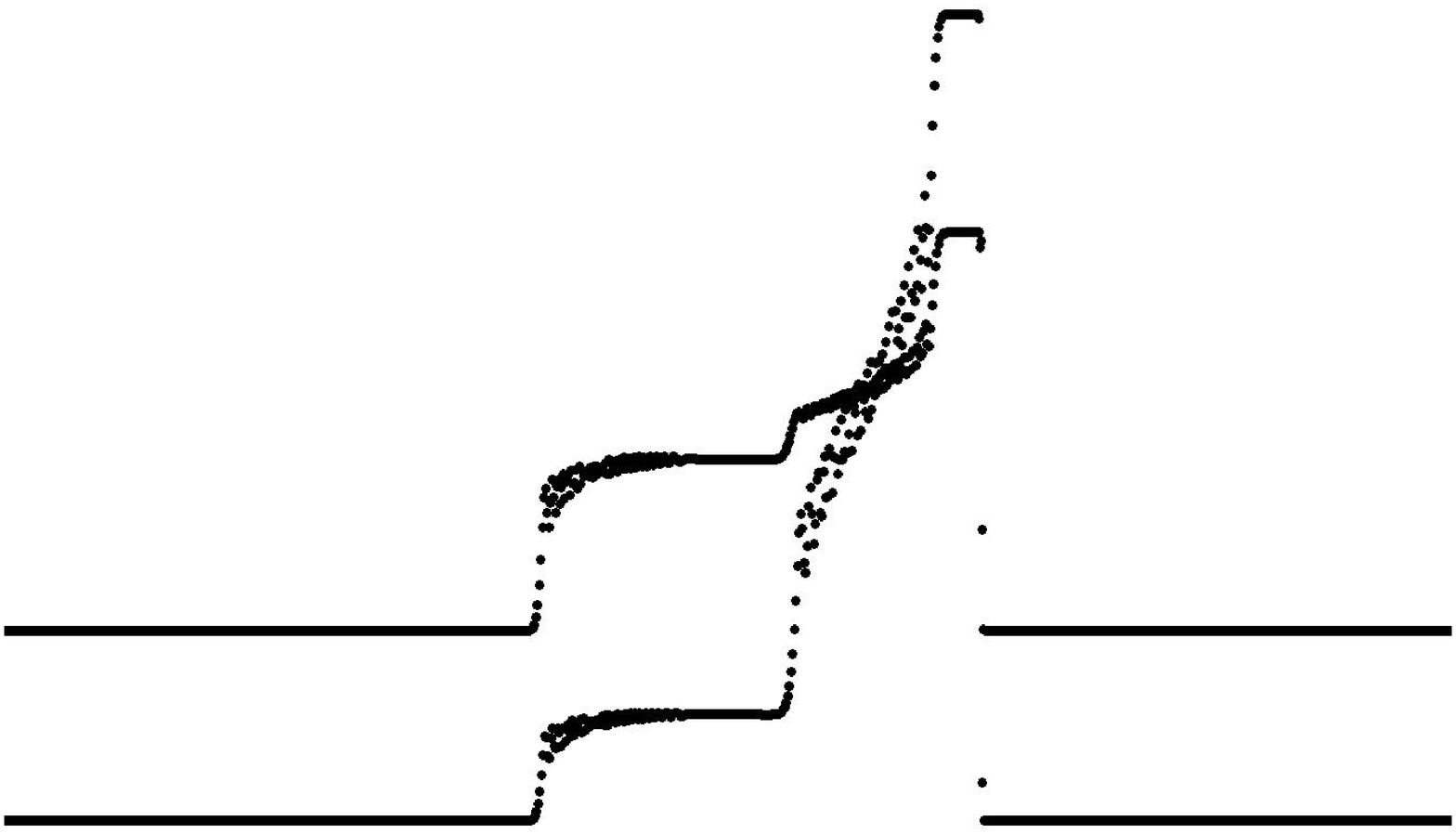}
        \caption{$t=250$}
        \label{fig1}
    \end{subfigure}
    \begin{subfigure}[b]{0.48\linewidth}        
        \centering
        \includegraphics[width=\linewidth]{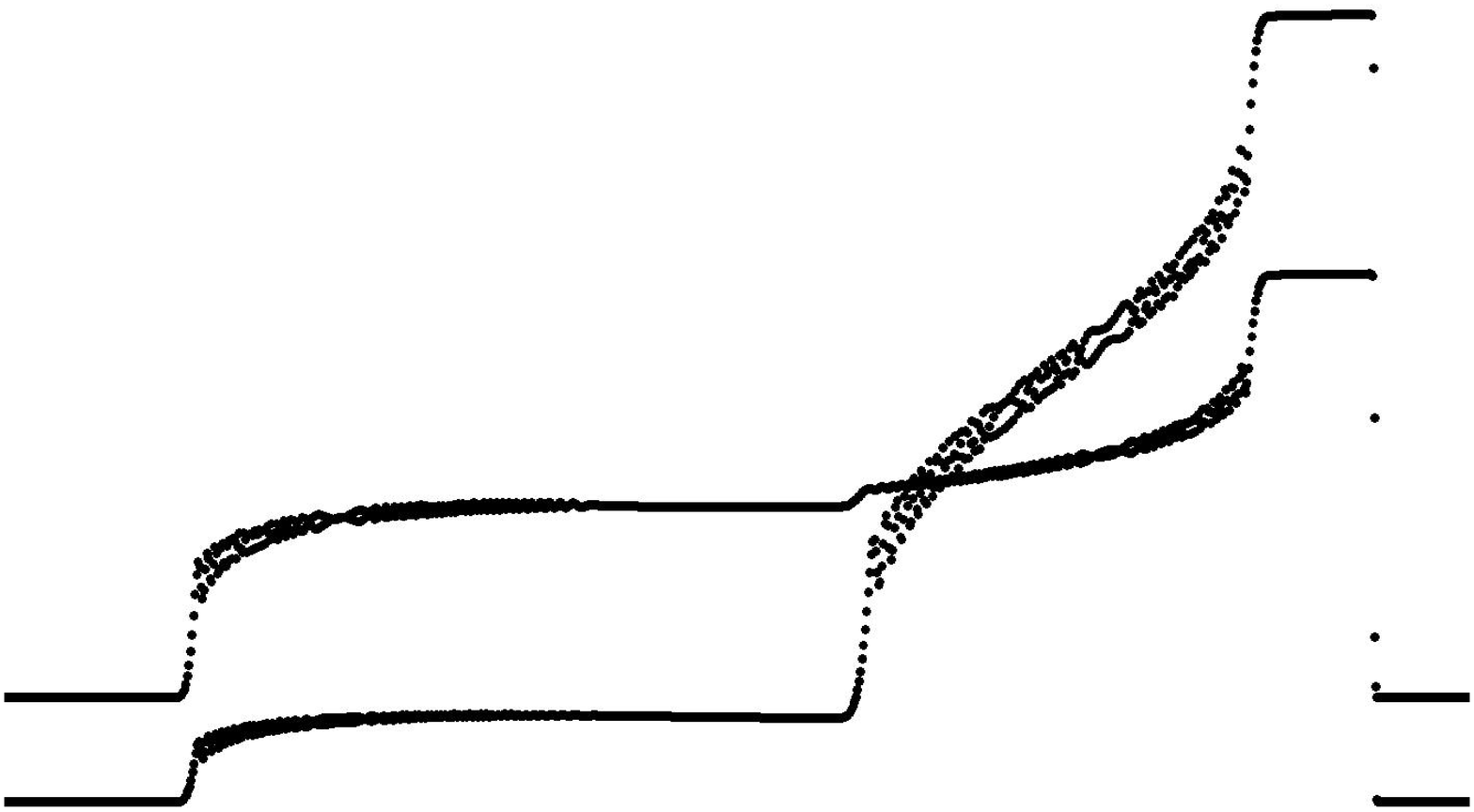}
        \caption{$t=700$}
        \label{fig2}
    \end{subfigure}
    \caption{Two snapshots at times $t=250$ (A) and $t=700$ (B) of the (numerically computed) 
    solution of a Toda lattice, with a period 2 initial condition.} 
    \label{fig12}
\end{figure}

Figure~\ref{fig12} exemplifies the general situation in the periodic background case.
As time goes to infinity, the (n,t) space is divided  into
several regions. There are three kinds of such regions:
there are regions of periodicity (the period being equal to the period of the unpertrubed lattice),
there are solitons in a periodic background, and then there are regions where the lattice undergoes  \say{modulated} oscillations
with a large (order $1/t$) frequency and slowly varying (with $n/t$) amplitude and phase. Phenomena appear in two different scales 
and are naturally expressed in two new variables: the \say{fast} one being $1/t$ and the \say{slow} one being $n/t$.
The regions of periodicity and the modulation regions are open cones bounded by half-lines (if we consider only positive times t)
emerging at the origin.
The soliton regions are small (in $1/t$)
regions around (some of) these half-lines. The slopes of the half-lines are the speeds of the solitons.

In each figure in Figure~\ref{fig12}, the two observed lines express the variables $x_n(t)$ 
as functions of the particle index $n$ at a frozen time $t$.
In some areas, the lines seem to be continuous. This is due to the fact that we have plotted
a huge number of particles (2048 particles) and also due to the 2-periodicity in space. So, one can think of
the two lines as the even- and odd-numbered particles of the lattice.

We first note  the single soliton which separates two regions of
apparent periodicity on the right. On soliton's left side, we observe three different
areas with apparently periodic solutions of  period two.
Finally, there are some transitional (modulation) regions
which interpolate between the different period two regions. 

A natural question is  whether this behaviour is ubiquitous  in any FPUT lattice. Namely, that the 
$(n,t)$ half-plane is divided by half-lines into
pure periodic and modulated regions as above, while sometimes solitons appear in the boundaries of such regions.

Standard KAM theory suggests that this might  happen only for small pertrubations while in general chaos can occur.
On the other hand our situation here is somewhat different to 
standard KAM problems in that we have non-periodic perturbations of a periodic 
lattice so the short range perturbations have \say{more space} to travel into.

\section{Simulations' Setup}

As mentioned in the first paragraph, we are dealing with one-dimensional, periodic, uniform, 
doubly infinite lattices with nearest neighbor interaction. So for our numerics, we will consider the ODE 
system (\ref{FPUT}) with $(n,t)\in\{1,2,\dots,N\}\times\R$ 
for fixed $N\in\mathbb{N}$ and impose the periodic condition $x_{N+1}=x_1$.
Defining $q_n:=x_n$ and $p_n:=\dot{q}_n=\dot{x}_n=\frac{dx_n}{dt}$, system (\ref{FPUT}) can be written as
\be
\begin{cases}
\dot{q}_n=p_n\\
\dot{p}_n=V'(q_{n+1}-q_n)-V'(q_n-q_{n-1}),\quad (n,t) \in\{1,2,\dots,N\}\times\R
\end{cases}
\ee
with Hamiltonian
\be
\mathcal{H}(\mathbf{q},\mathbf{p})=\sum_{n=1}^{N}\Bigg[\frac{1}{2}p_n^2+V(q_{n+1}-q_n)\Bigg]
\ee
where $\mathbf{q}=(q_1,q_2,\dots,q_N)$ and $\mathbf{p}=(p_1,p_2,\dots,p_N)$. As far as the initial
conditions are concerned, we either require 
\begin{itemize}
\item
perturbed zero (or trivial) background conditions, specifically 
\be
\begin{cases}
q_n(0)=exp\{-\big(\frac{n-\frac{N}{4}}{4}\big)^2\}\\
p_n(0)=0
\end{cases}, n=1,2,\dots,N\quad\text{or}
\ee
\item
perturbed periodic background conditions (the period being $2$), i.e.
\be
\begin{cases}
q_n(0)=0\\
p_n(0)=(-1)^n+2\delta_{n}^{N/2}
\end{cases}, n=1,2,\dots,N
\ee
where $\delta_i^j$ denotes Kronecker's delta.
\end{itemize} 

Our simulations are based on MATLAB\textsuperscript{\textregistered} in which we consider $N=2048$ and use
$ode45$ as an integration method. For the time discretization we use a time-step size of $1$ for a total number of $800$ steps. 
The algorithm (see appendix \ref{code}) is similar to that found in Scholarpedia's article about the 
\href{http://www.scholarpedia.org/article/Fermi-Pasta-Ulam_nonlinear_lattice_oscillations}{FPUT nonlinear lattice oscillations}
which in fact comes from \cite{dpr}. Finally, for the potential function $V$ we consider the following candidates
\begin{itemize}
\item
FPUT potential $V(x)= \frac{1}{2} x^2 + \frac{\alpha}{3} x^3 + \frac{\beta}{4} x^4$, where 
$\alpha,\beta$ are real parameters. More presicely we only consider its two offsprings, the
FPUT$-\alpha$ potential (for $\beta=0$) $V(x)= \frac{1}{2} x^2 + \frac{\alpha}{3} x^3$ and the
FPUT$-\beta$ potential (for $\alpha=0$) $V(x)= \frac{1}{2} x^2 + \frac{\beta}{4} x^4$
\item
harmonic potential $V(x)=\frac{1}{2}x^2$
\item
Hertz potential 
$V(x)=
\begin{cases}
c|x|^{5/2},\quad x<0\\
0,\quad x\geq0
\end{cases}$, where $c$ is a real parameter
\item
Langmuir (or Volterra or Kac-van Moerbeke or Moser or discrete KdV) potential $V(x)=e^{x}$
\item
perturbed Langmuir potential $V(x)=e^{x}+\alpha x^3+\beta x^4$, where $\alpha,\beta$ are real parameters.
We study the cases $V(x)=e^{x}+\alpha x^3$ and $V(x)=e^{x}+\beta x^4$ separately.
\item
$(2,1)$Lennard-Jones potential $V(x)=\varepsilon\bigg[\Big(\frac{d}{d+x}\Big)^2-2\frac{d}{d+x}+1\bigg]$, 
where $\varepsilon, d$ are real parameters
\item
Morse potential $V(x)=\gamma(e^{-\delta x}-1)^2$, where $\gamma,\delta$ are real parameters
\item
Toda potential $V(x)=e^{-x}+x$
\item
perturbed Toda potential $V(x)=e^{-x}+x+\alpha x^3+\beta x^4$, where $\alpha,\beta$ are real parameters.
We study the cases $V(x)=e^{-x}+x+\alpha x^3$ and $V(x)=e^{-x}+x+\beta x^4$ separately.
\end{itemize}

Closing this paragraph, it is essential to add that all of our numerics have been checked for accuracy in the sense that
the quantities (e.g. total momentum, hamiltonian) that are expected to be conserved are indeed 
(almost) conserved. We observed only very small deviations from these constant values.

\section{Numerical results with trivial background}

In this section we present some numerical experiments which support the soliton resolution conjecture
in the case of a lattice with trivial background. Here, and in the next section, 
we plot $q_n$ as a function of $n$ at two specific times.
Again,  $n$ is a discrete variable, but our pictures cover around 1800 particles (excluding 124 from each side
of the altogether 2048 particle chain), so what should be a 
discrete sequence of dots may look like a smooth curve. If we zoomed in, we should be able to distinguish the dots.
Again, in the integrable cases (i.e. Langmuir and Toda), the result can be proved (\cite{nh}, \cite{k1}, \cite{krt2}) 
with the help of the inverse scattering theory.

\begin{figure}[H]
    \centering
    \begin{subfigure}[b]{0.48\linewidth}        
        \centering
        \includegraphics[width=\linewidth]{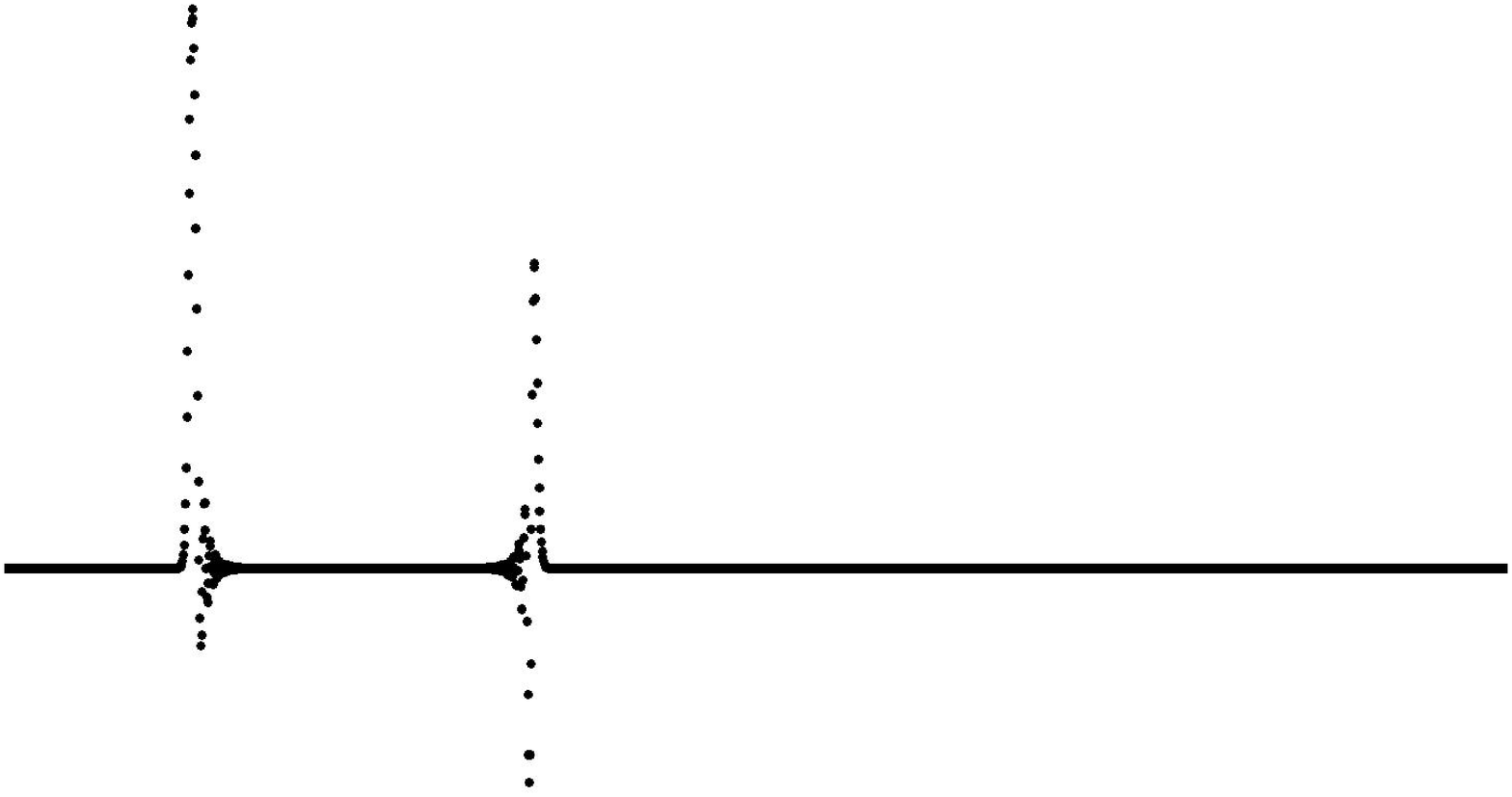}
        \caption{$t=225$}
        \label{fig3}
    \end{subfigure}
    \begin{subfigure}[b]{0.48\linewidth}        
        \centering
        \includegraphics[width=\linewidth]{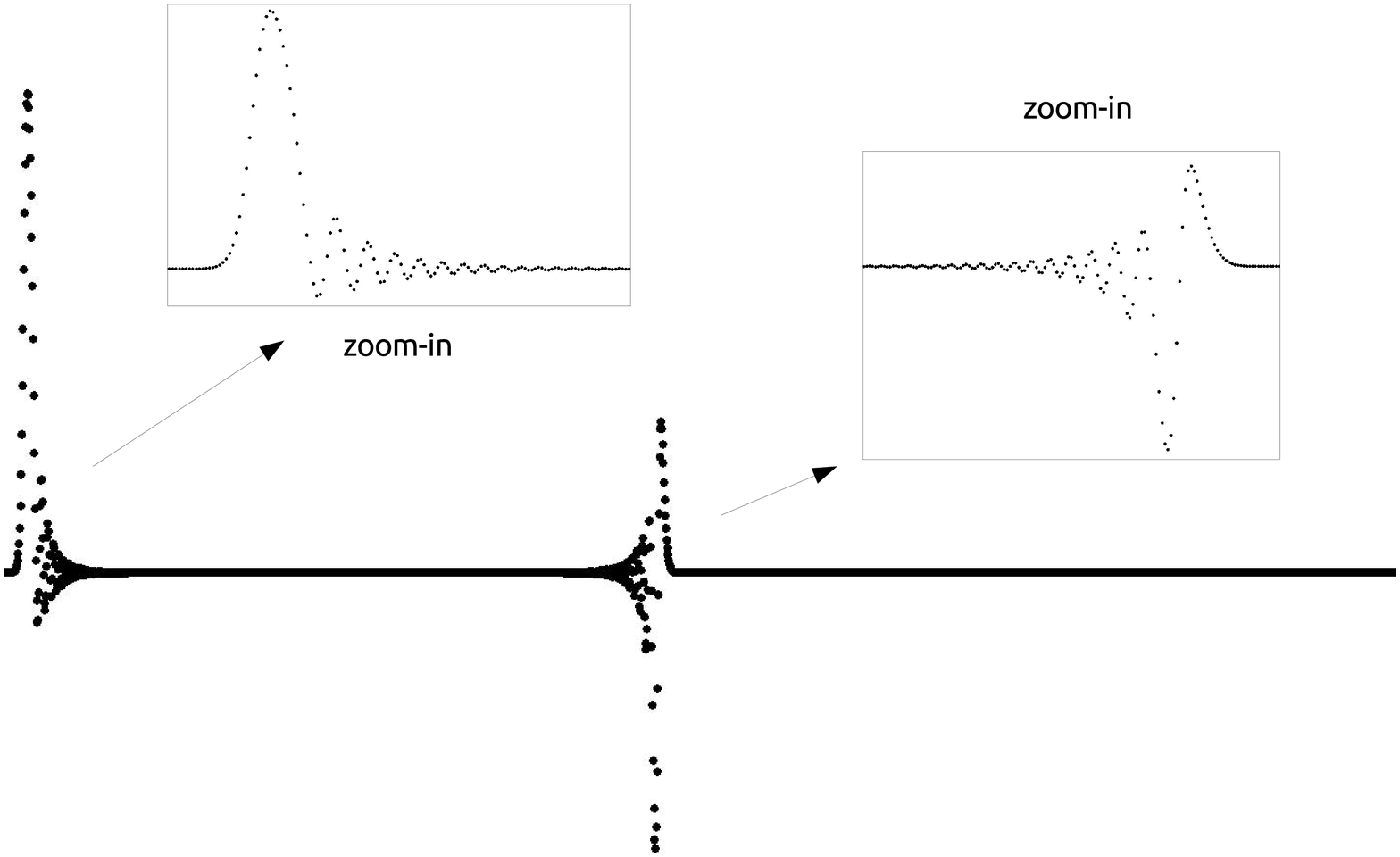}
        \caption{$t=450$}
        \label{fig4}
    \end{subfigure}
\caption{Two snapshots at times $t=225$ (A) and $t=450$ (B) of the (numerically computed) 
         solution of a FPUT$-\alpha$ lattice for $\alpha=0.25$ with zero background initial condition.} 
    \label{fig34}
\end{figure}

Our first numerical simulation is concerned with the FPUT lattice. More specifically with FPUT$-\alpha$
potential. We have completed experiments with different values of the $\alpha$ parameter.
We put $\alpha=0.425$, $0.4$, $0.25$, $0.1$ and $0.01$. All results turned out to be qualitively the same.
Figure \ref{fig34} shows these results in the case of an FPUT-$\alpha$ potential with $\alpha=0.25$. 

\begin{figure}[H]
    \centering
    \begin{subfigure}[b]{0.48\linewidth}        
        \centering
        \includegraphics[width=\linewidth]{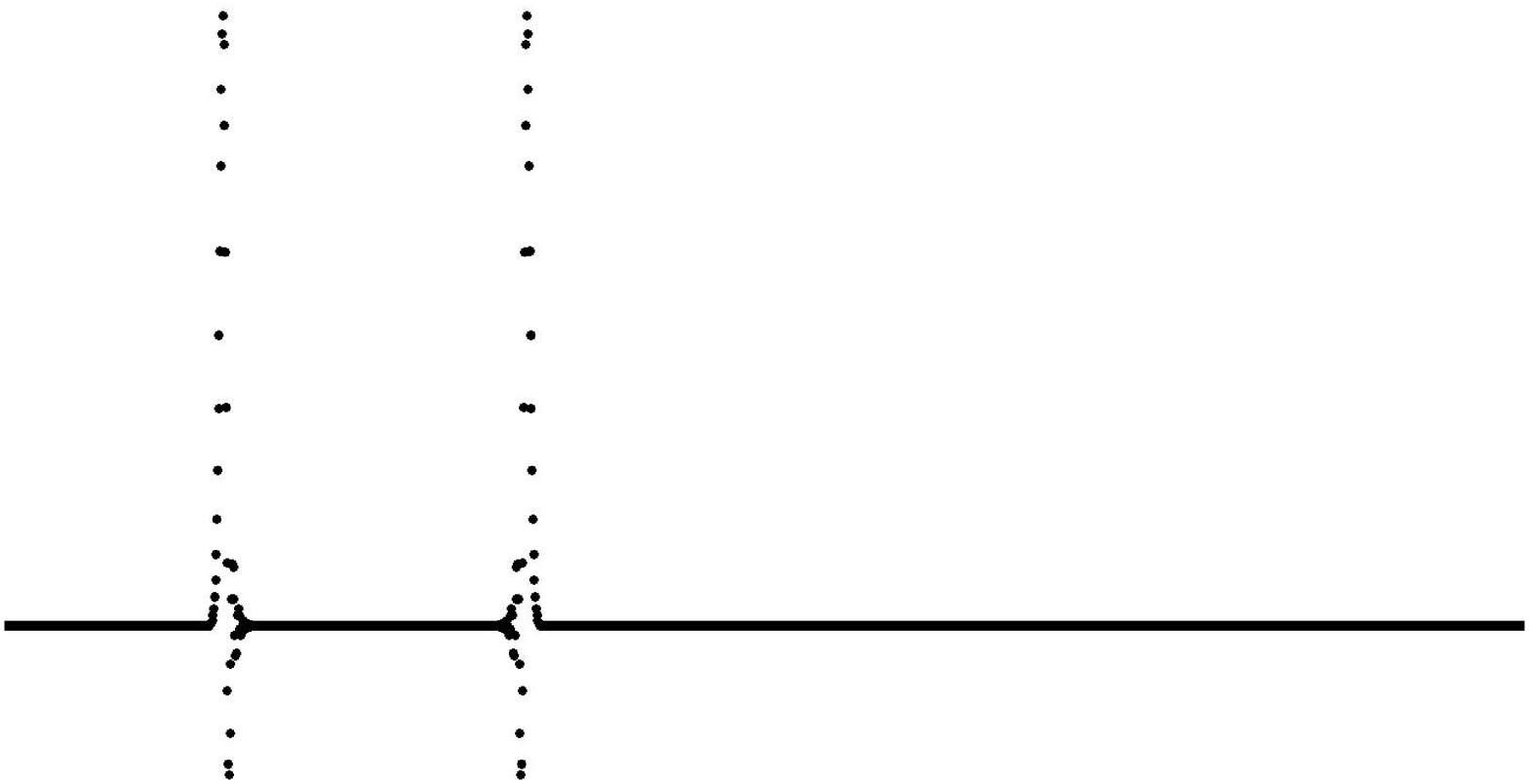}
        \caption{$t=200$}
        \label{fig5}
    \end{subfigure}
    \begin{subfigure}[b]{0.48\linewidth}        
        \centering
        \includegraphics[width=\linewidth]{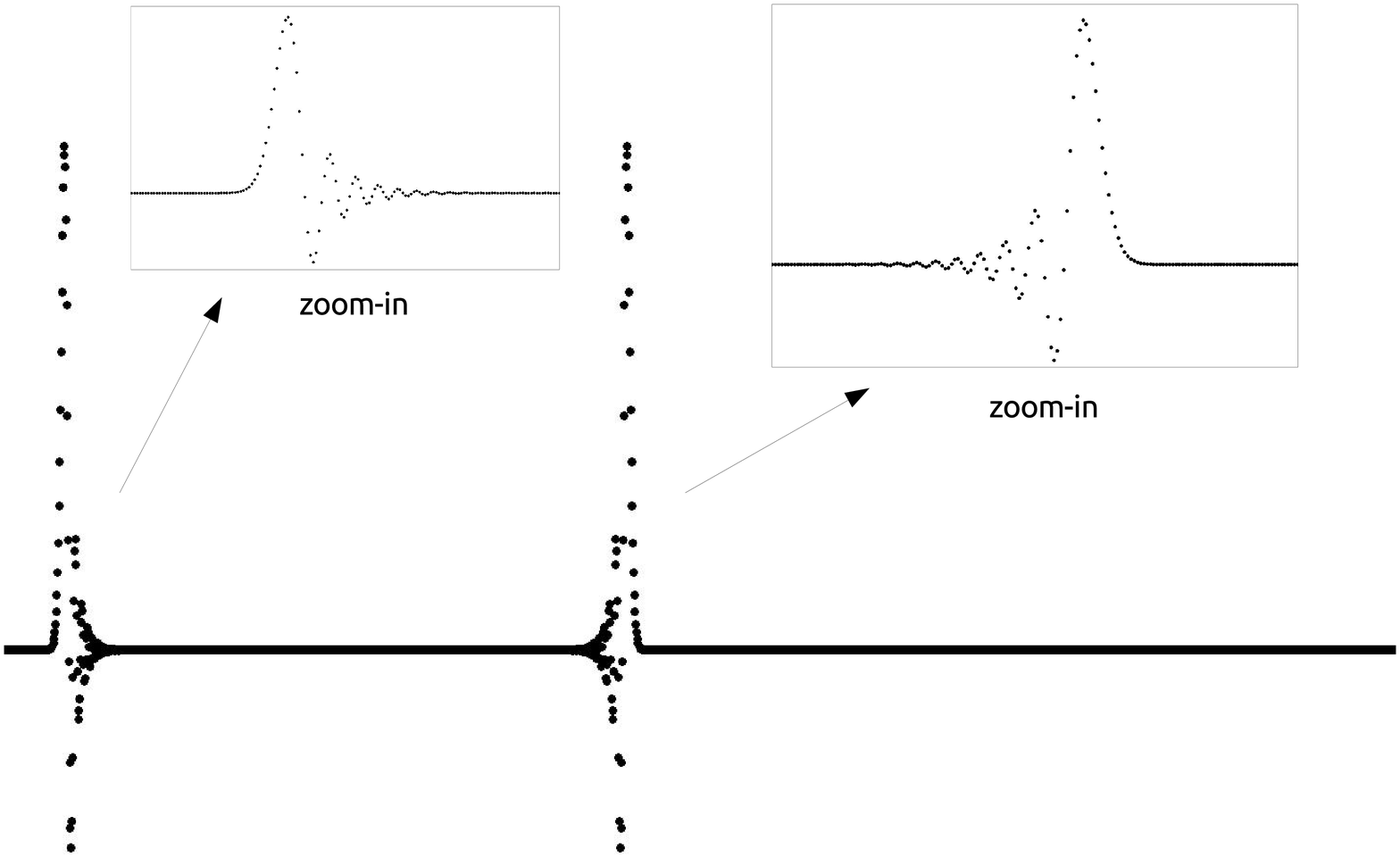}
        \caption{$t=400$}
        \label{fig6}
    \end{subfigure}
\caption{Two snapshots at times $t=200$ (A) and $t=400$ (B) of the (numerically computed) 
         solution of a FPUT$-\beta$ lattice for $\beta=0.01$ with zero background initial condition.}
    \label{fig56}
\end{figure}

Next, we experimented with the FPUT-$\beta$ potential. As before, we tried $\alpha=0.425$, $0.4$, $0.25$, 
$0.1$ and $0.01$. Once more, all the outcomes had the same qualitative nature. In Figure \ref{fig56} we 
present the results of our numerics for the FPUT-$\beta$ potential with $\beta=0.01$. 

In both cases the soliton resolution is crystal clear! We observe two well-defined solitons with constant 
amplitude and shape and well defined constant speeds. The background radiation is very small.
We have conducted many more simulations for the case of zero background.
Following is a list of the potentials that gave similar results (identical pictures with the figures above)
\begin{itemize}
\item
harmonic
\item
Langmuir
\item
small perturbations (e.g. $\alpha,\beta=0.01$ or less) of the Langmuir 
\item
$(2,1)-$Lennard-Jones for \say{big} values of the parameter $d$ representing lattice spacing
(in equilibrium), e.g. $d=10$ or larger. In this case, $\varepsilon$ can be anything
\item
Morse for \say{small} values of the parameter $\delta$. $\gamma$ takes arbitrary values.
\item
Toda 
\item
small perturbations (e.g. $\alpha,\beta=0.1$ or less) of Toda 
\end{itemize} 

On the other hand, other  experiments give  something different! The following pictures
show a representative sample of them.
\begin{figure}[H]
    \centering
    \begin{subfigure}[b]{0.48\linewidth}        
        \centering
        \includegraphics[width=\linewidth]{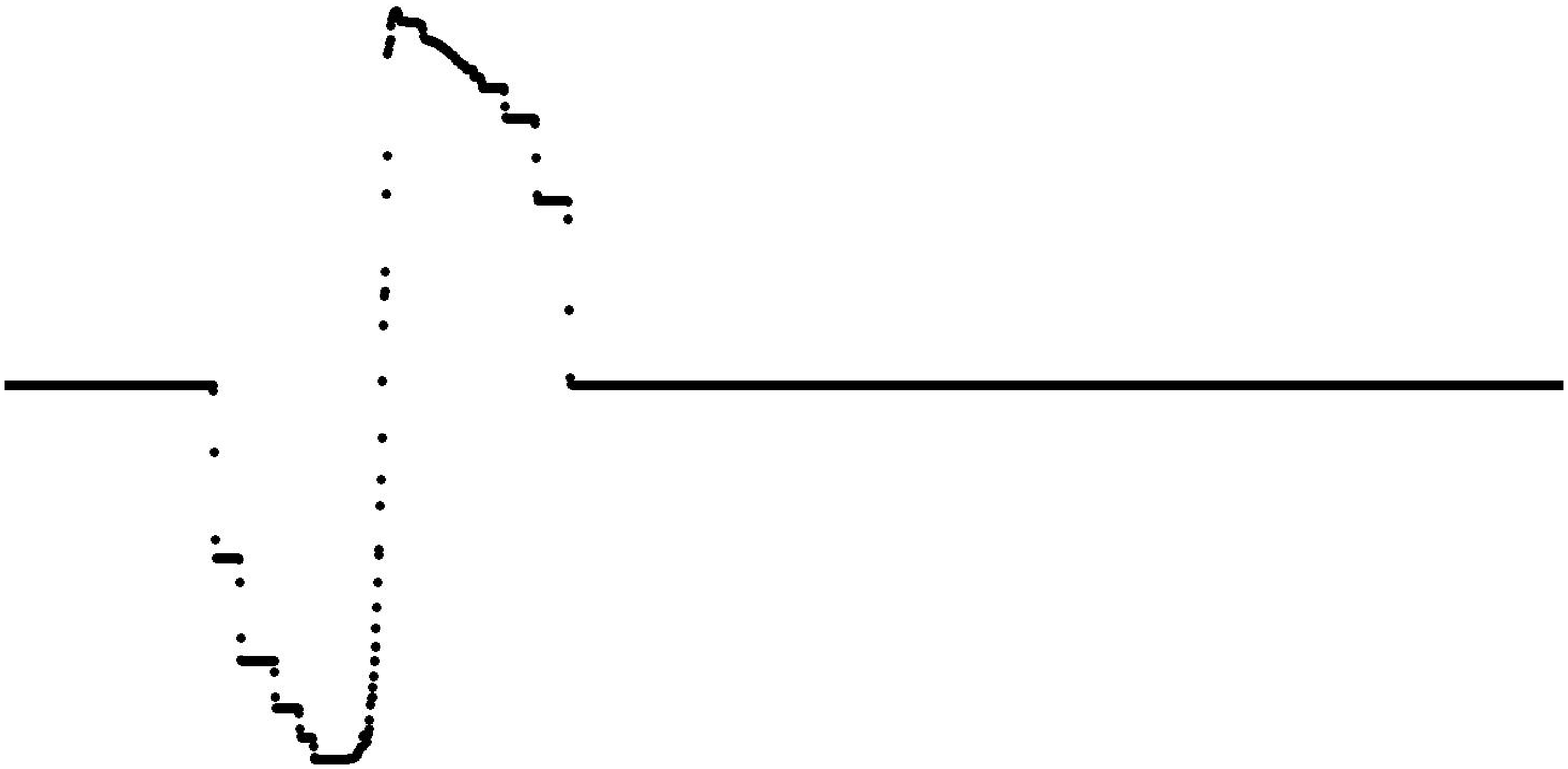}
        \caption{$t=250$}
    \end{subfigure}
    \begin{subfigure}[b]{0.48\linewidth}        
        \centering
        \includegraphics[width=\linewidth]{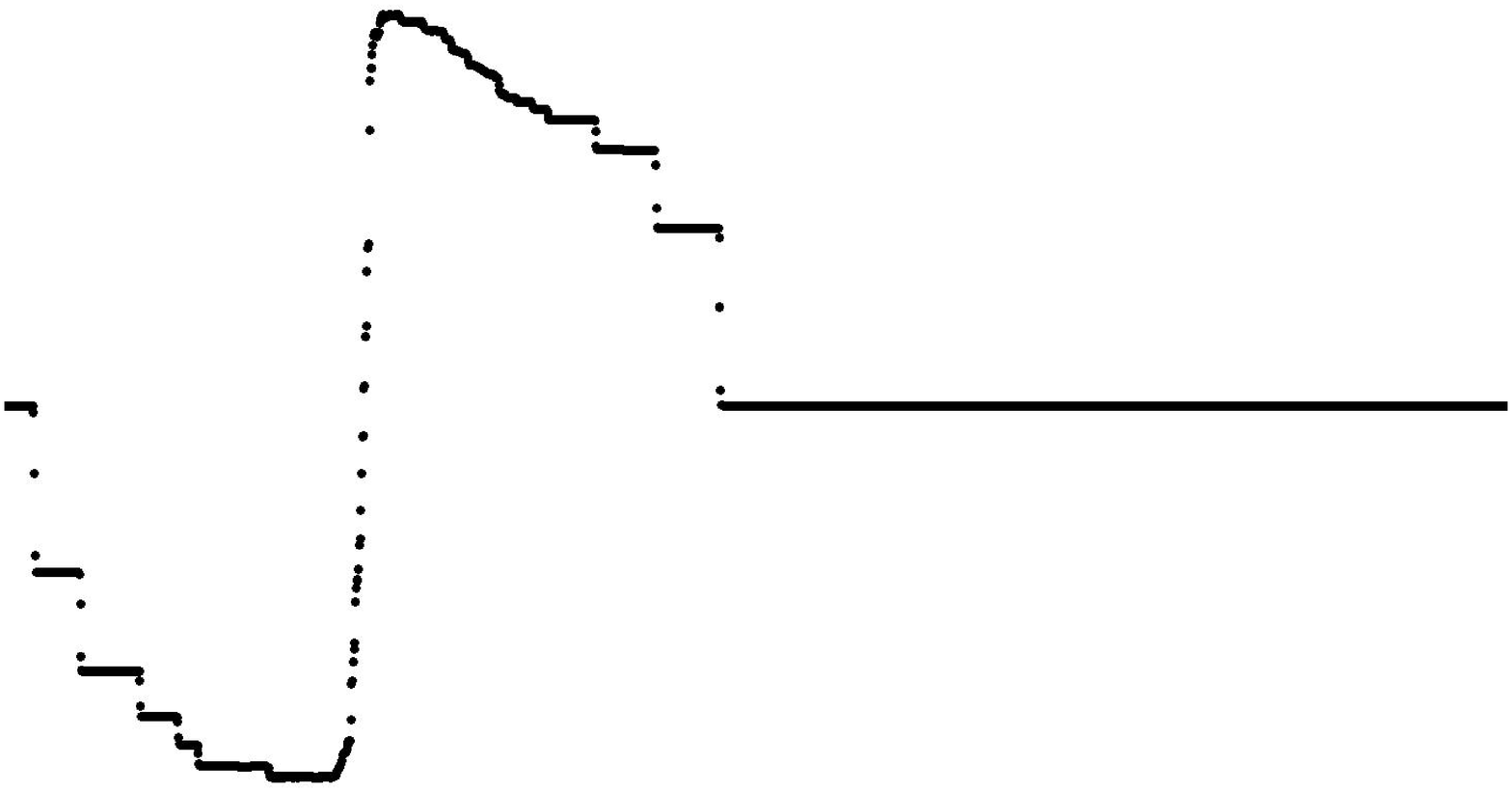}
        \caption{$t=500$}
    \end{subfigure}
\caption{Two snapshots at times $t=250$ (A) and $t=500$ (B) of the (numerically computed) 
         solution of a Hertz lattice for $c=1$ with zero background initial condition.}
\end{figure}

\begin{figure}[H]
    \centering
    \begin{subfigure}[b]{0.48\linewidth}        
        \centering
        \includegraphics[width=\linewidth]{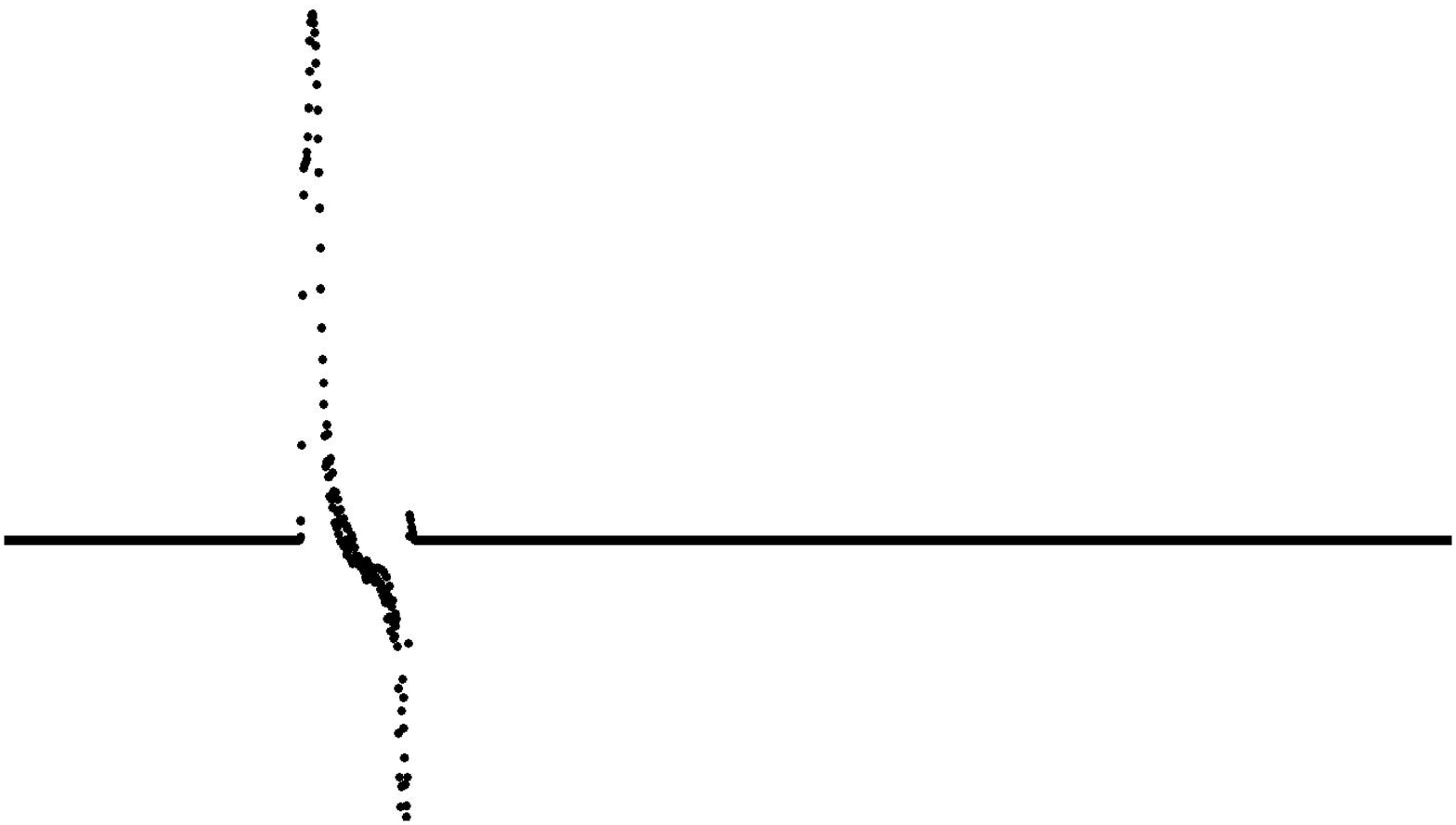}
        \caption{$t=200$}
    \end{subfigure}
    \begin{subfigure}[b]{0.48\linewidth}        
        \centering
        \includegraphics[width=\linewidth]{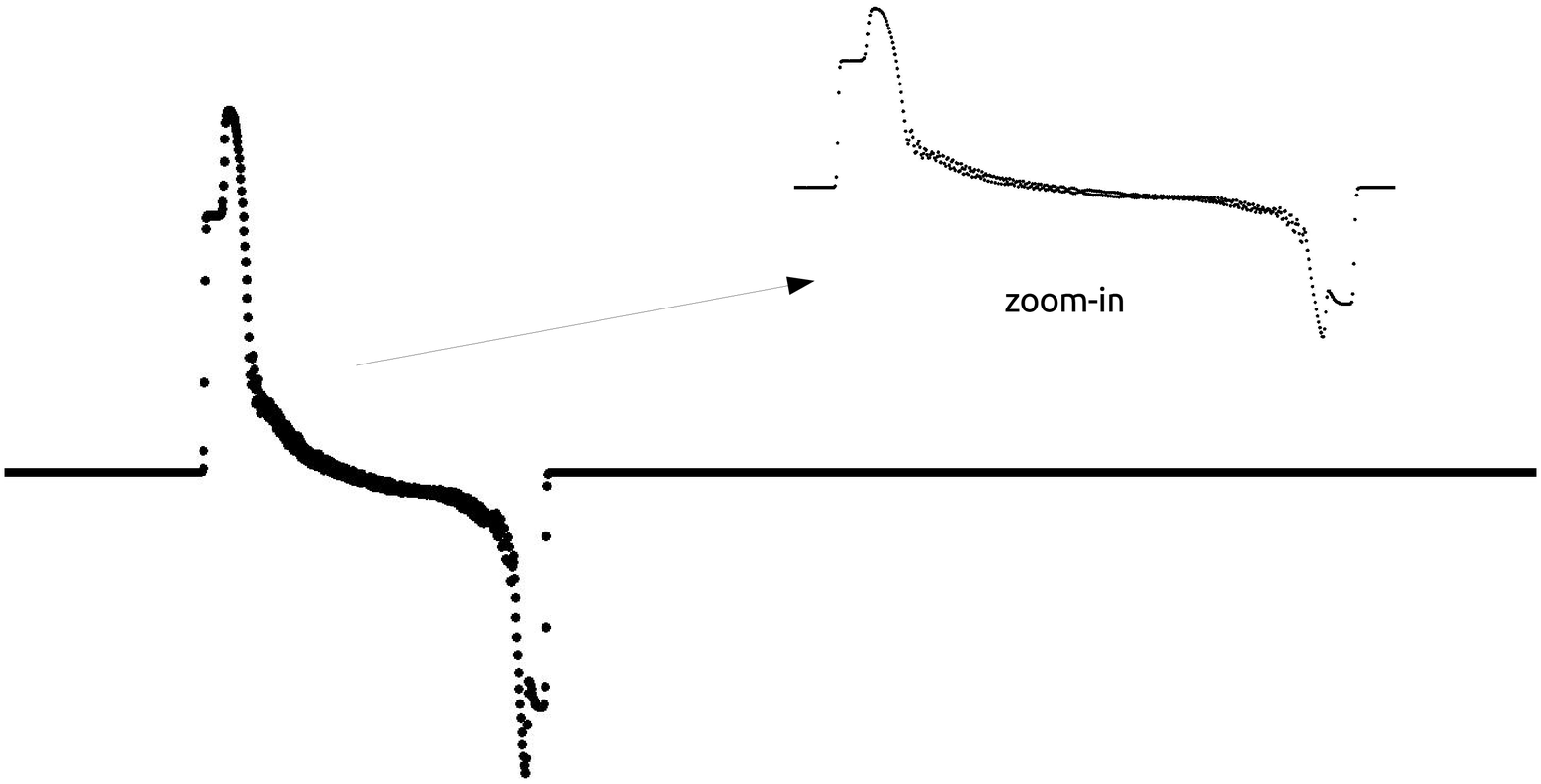}
        \caption{$t=600$}
    \end{subfigure}
\caption{Two snapshots at times $t=200$ (A) and $t=600$ (B) of the (numerically computed) 
         solution of a Langmuir lattice with a cubic perturbation of $\alpha=0.1$ with zero 
         background initial condition.}
\end{figure}

Below, in Figure \ref{m+lj} we see the results coming from the integrator for
a $(2,1)-$Lennard-Jones lattice for $d=1$ and $\varepsilon=10$ with zero background initial condition.
Qualitatively, we get the same result for a Morse potential with $\gamma=1/2$ and $\delta=1$.

\begin{figure}[H]
    \centering
    \begin{subfigure}[b]{0.48\linewidth}        
        \centering
        \includegraphics[width=\linewidth]{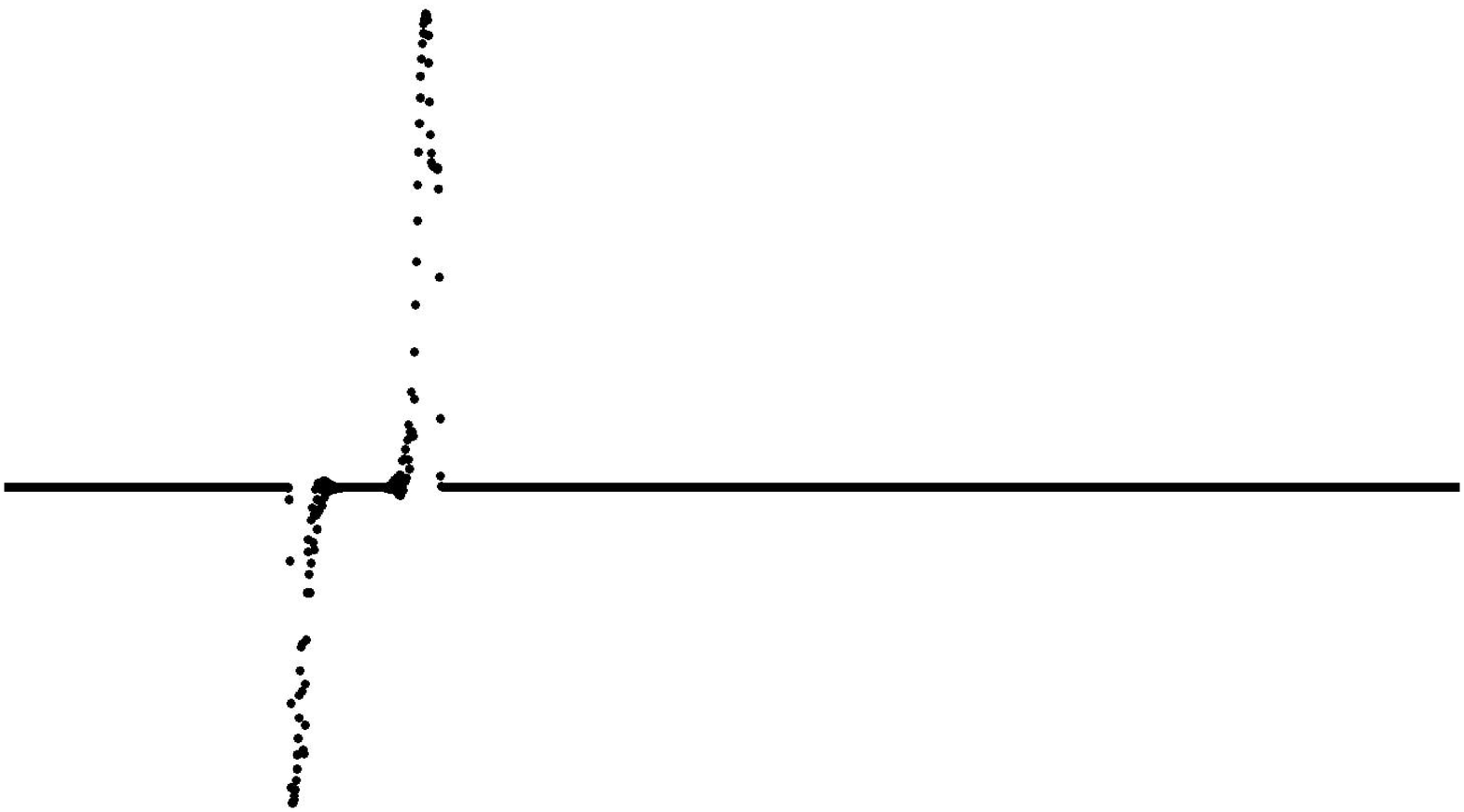}
        \caption{$t=20$}
    \end{subfigure}
    \begin{subfigure}[b]{0.48\linewidth}        
        \centering
        \includegraphics[width=\linewidth]{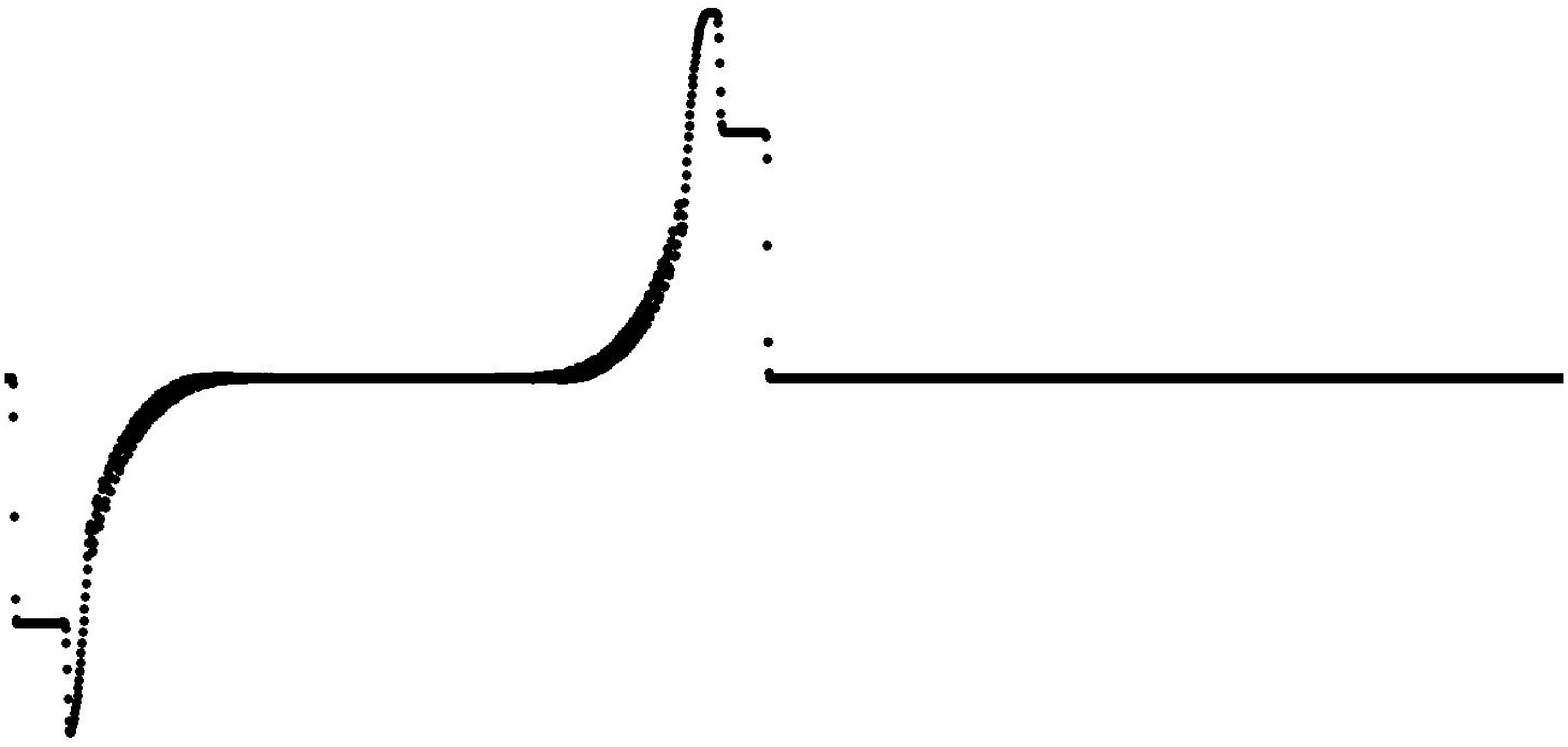}
        \caption{$t=90$}
    \end{subfigure}
\caption{Two snapshots at times $t=20$ (A) and $t=90$ (B) of the (numerically computed) 
         solution of a $(2,1)-$Lennard-Jones lattice for $d=1$ and $\varepsilon=10$ with zero 
         background initial condition.}
         \label{m+lj}
\end{figure}

\begin{figure}[H]
    \centering
    \begin{subfigure}[b]{0.48\linewidth}        
        \centering
        \includegraphics[width=\linewidth]{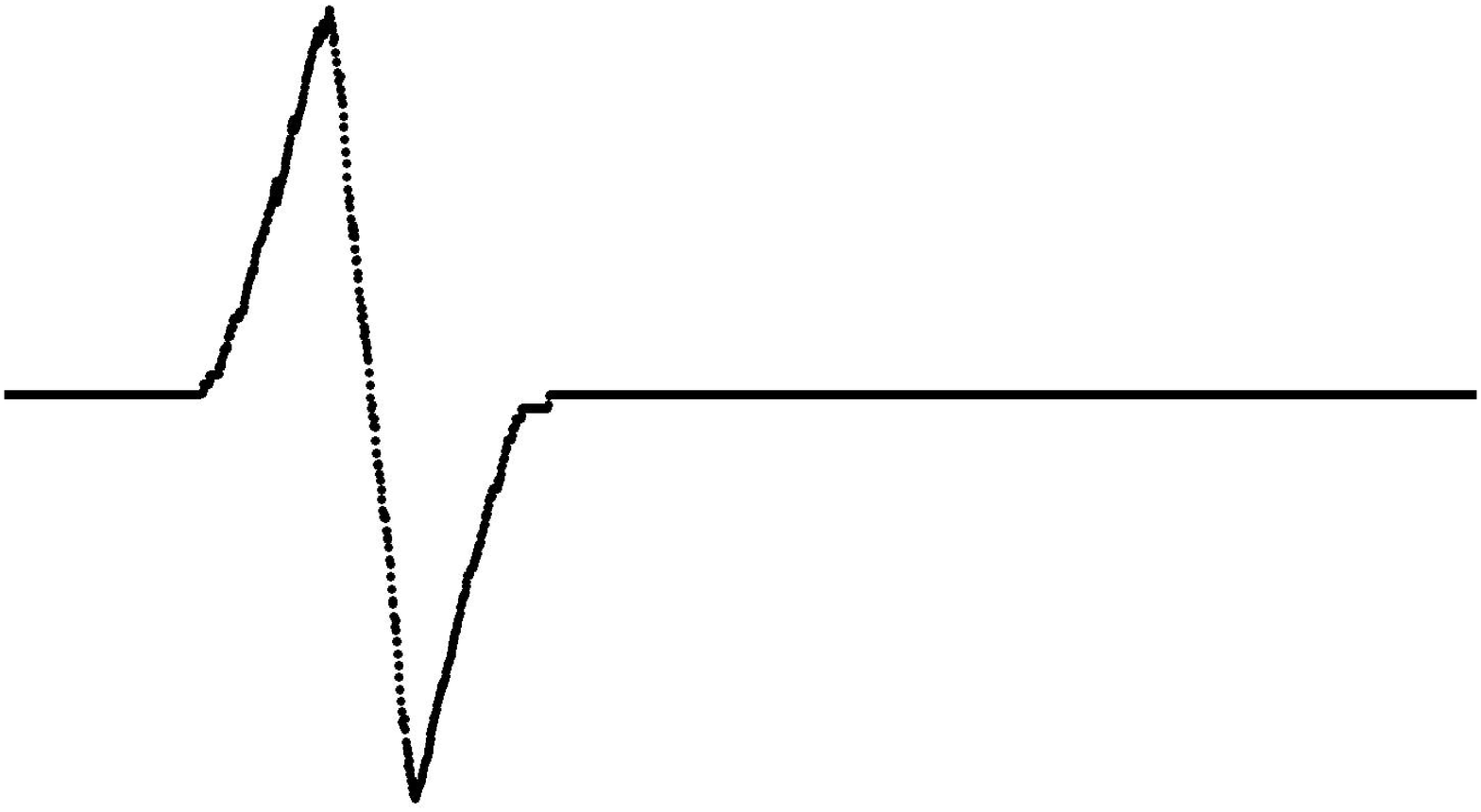}
        \caption{$t=25$}
    \end{subfigure}
    \begin{subfigure}[b]{0.48\linewidth}        
        \centering
        \includegraphics[width=\linewidth]{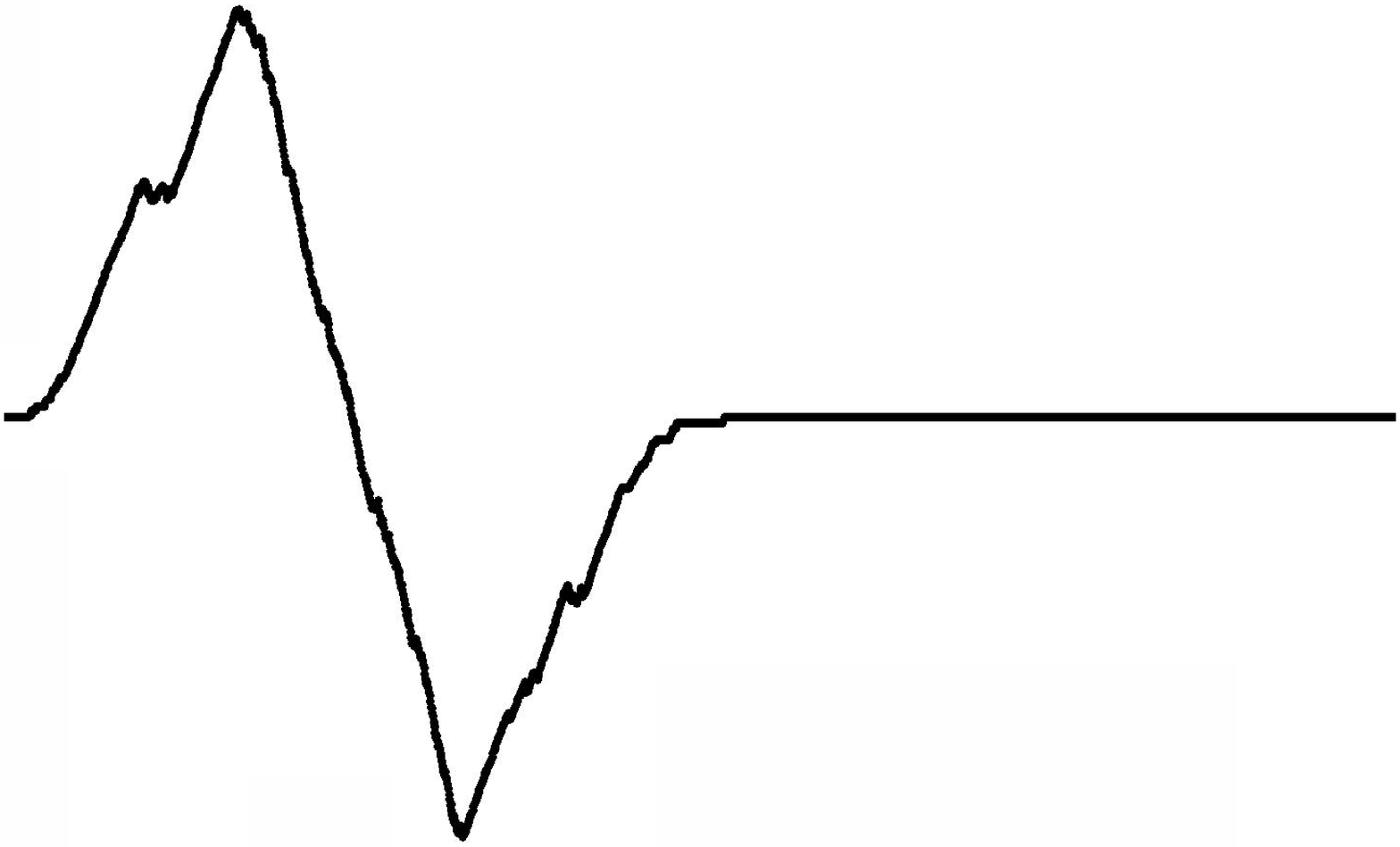}
        \caption{$t=50$}
    \end{subfigure}
\caption{Two snapshots at times $t=25$ (A) and $t=50$ (B) of the (numerically computed) 
         solution of a Toda lattice with a cubic perturbation of $\alpha=10$ and zero 
         background initial condition.}
\end{figure}

\begin{figure}[H]
    \centering
    \begin{subfigure}[b]{0.48\linewidth}        
        \centering
        \includegraphics[width=\linewidth]{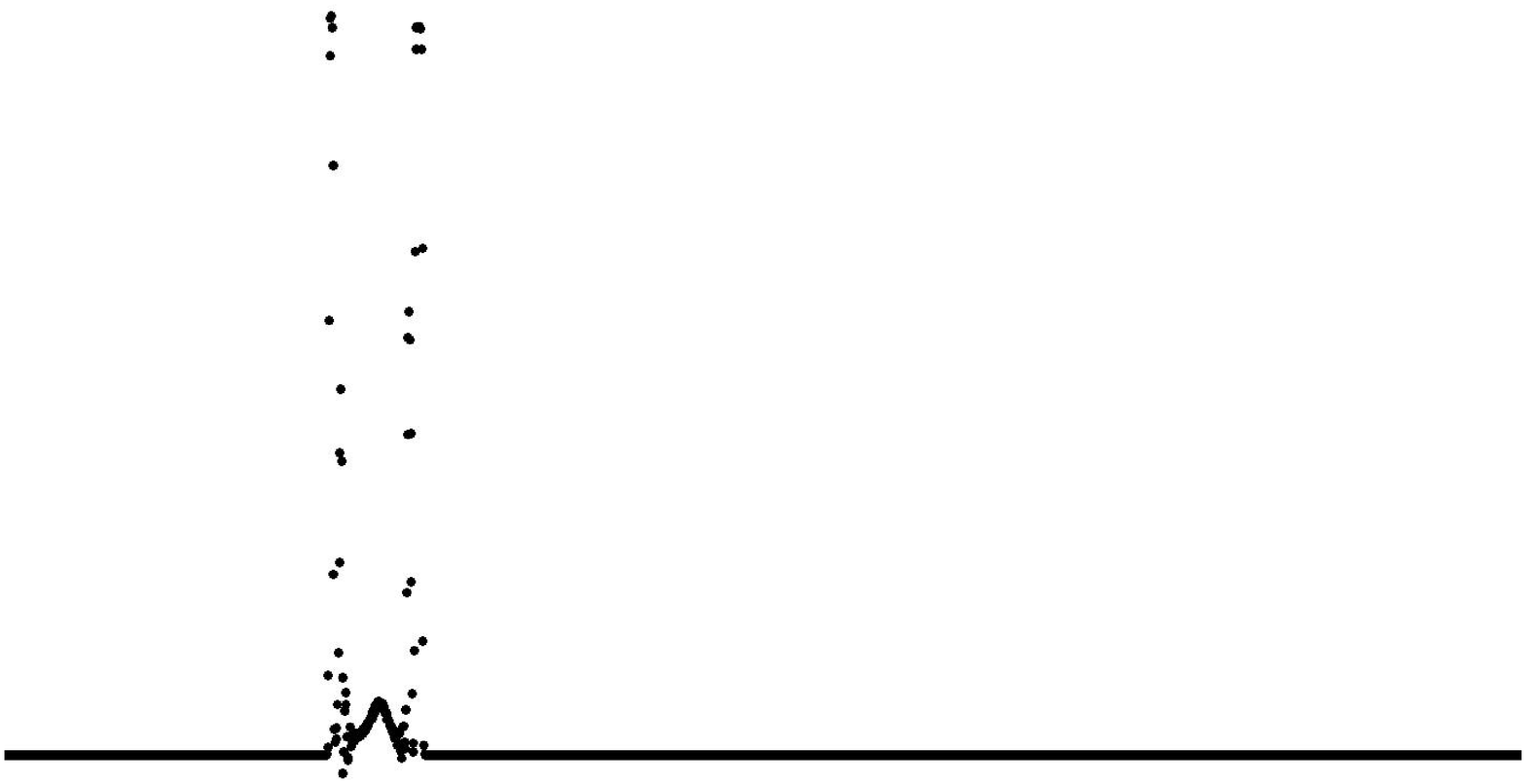}
        \caption{$t=25$}
    \end{subfigure}
    \begin{subfigure}[b]{0.48\linewidth}        
        \centering
        \includegraphics[width=\linewidth]{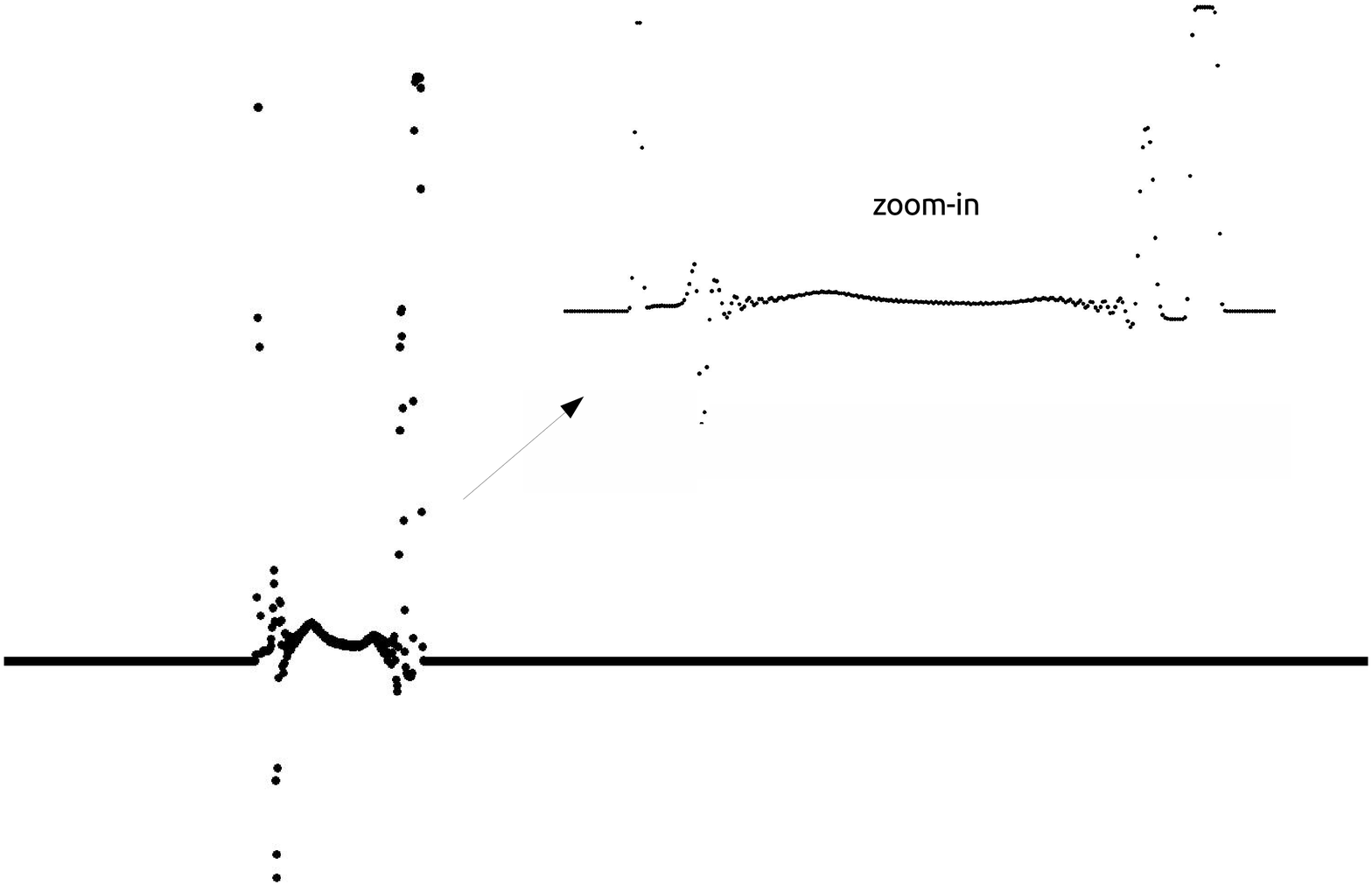}
        \caption{$t=50$}
    \end{subfigure}
\caption{Two snapshots at times $t=25$ (A) and $t=50$ (B) of the (numerically computed) 
         solution of a Toda lattice with a quartic perturbation of $\beta=10$ and zero 
         background initial condition.}
\end{figure}

\section{Numerical results with periodic background}

In this section we present some numerical experiments which support our amended soliton resolution 
conjecture in the case of a lattice with nearest neighbour interaction but in this case for a periodic 
background. In the case of the harmonic lattice, one has the following findings
\begin{figure}[H]
    \centering
    \begin{subfigure}[b]{0.48\linewidth}        
        \centering
        \includegraphics[width=\linewidth]{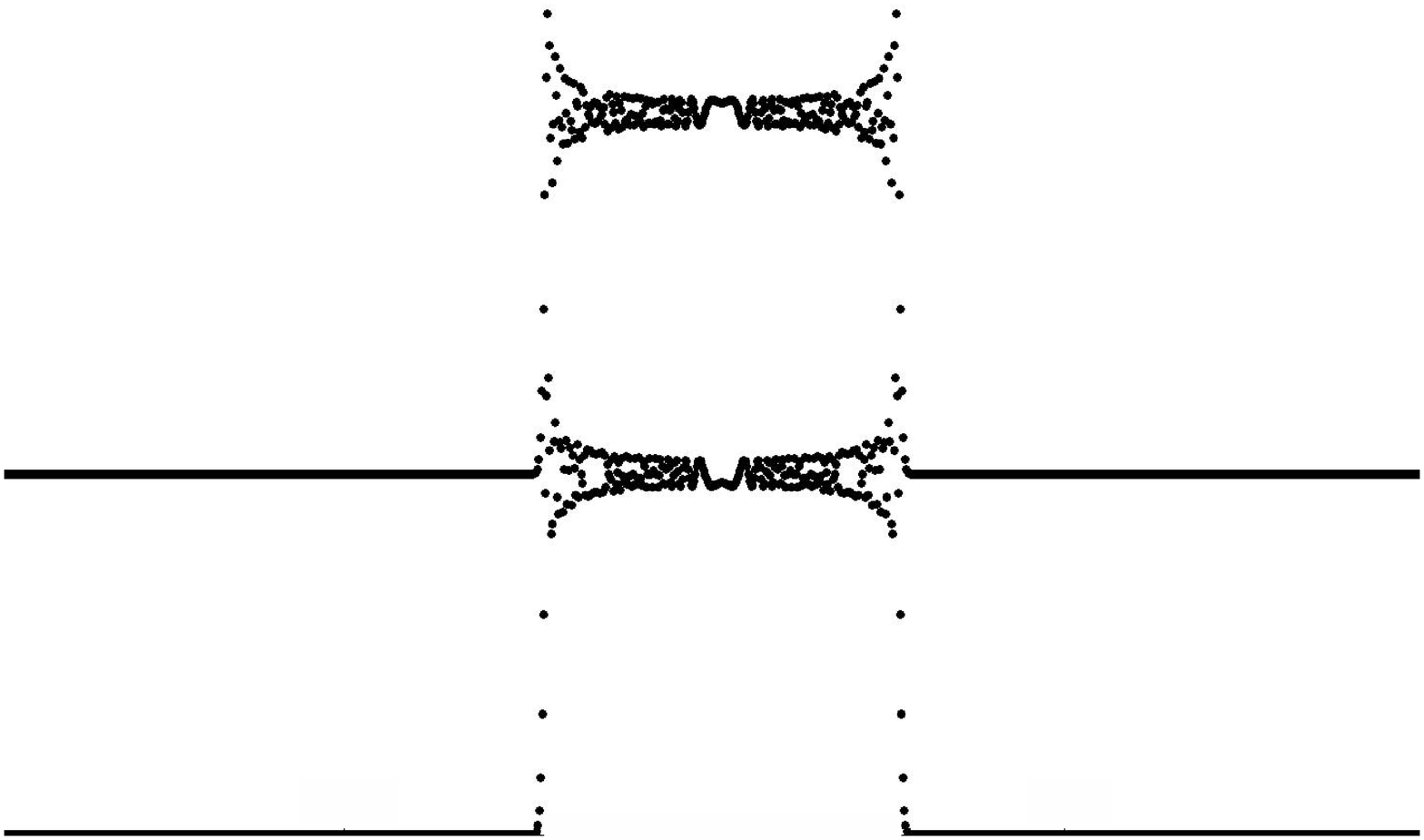}
        \caption{$t=250$}
    \end{subfigure}
    \begin{subfigure}[b]{0.48\linewidth}        
        \centering
        \includegraphics[width=\linewidth]{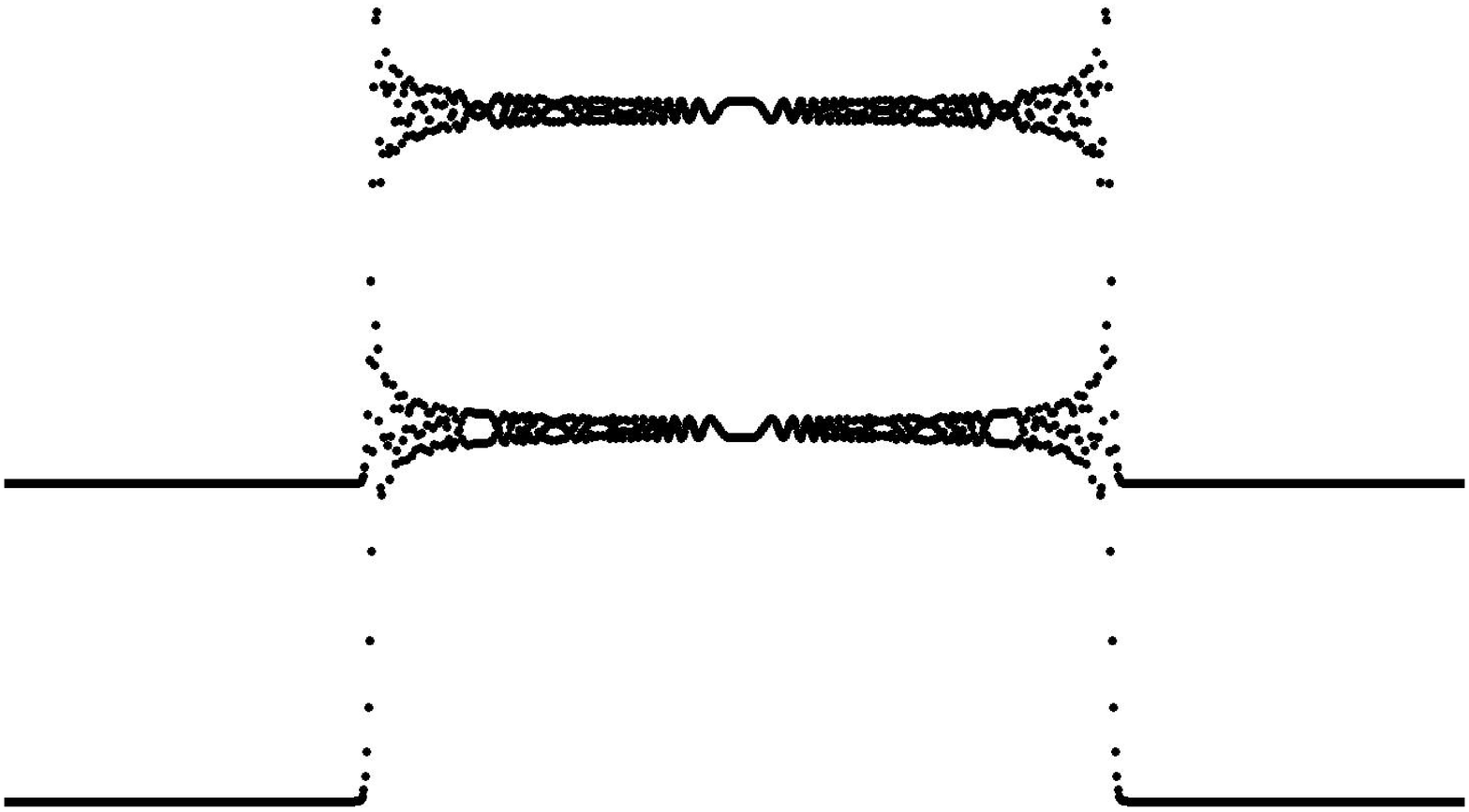}
        \caption{$t=500$}
    \end{subfigure}
\caption{Two snapshots at times $t=250$ (A) and $t=500$ (B) of the (numerically computed) 
         solution of a harmonic lattice with periodic background initial condition.}
\end{figure}

Next, we continue with some pictures of FPUT$-\alpha$ and $\beta$ potentials and small values of these 
parameters (it can be said that these constitute \say{small} perturbations of the linear harmonic lattice). 
More precisely, for the FPUT$-\alpha$ potential and for $\alpha=0.25$, we have
\begin{figure}[H]
    \centering
    \begin{subfigure}[b]{0.48\linewidth}        
        \centering
        \includegraphics[width=\linewidth]{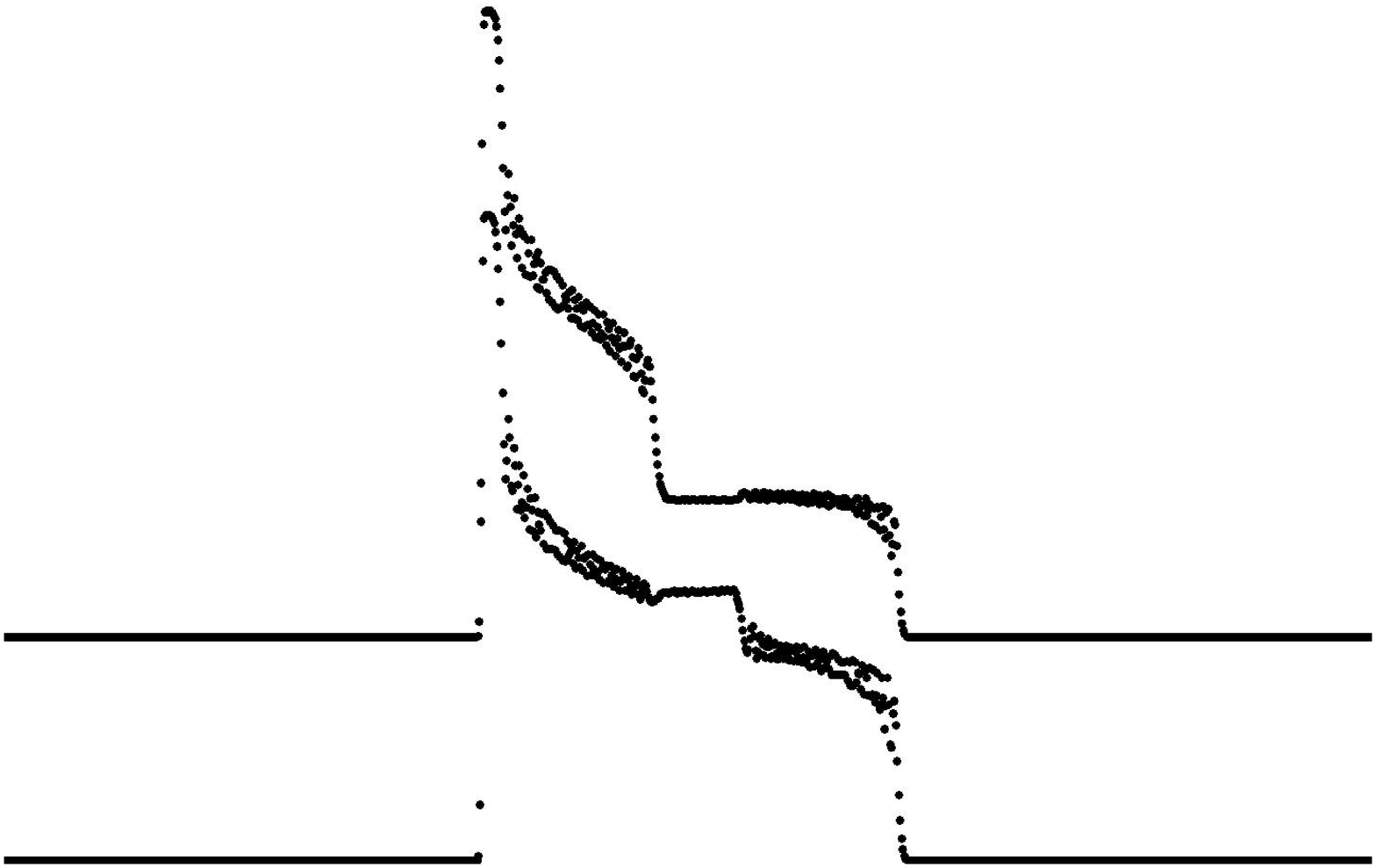}
        \caption{$t=300$}
    \end{subfigure}
    \begin{subfigure}[b]{0.48\linewidth}        
        \centering
        \includegraphics[width=\linewidth]{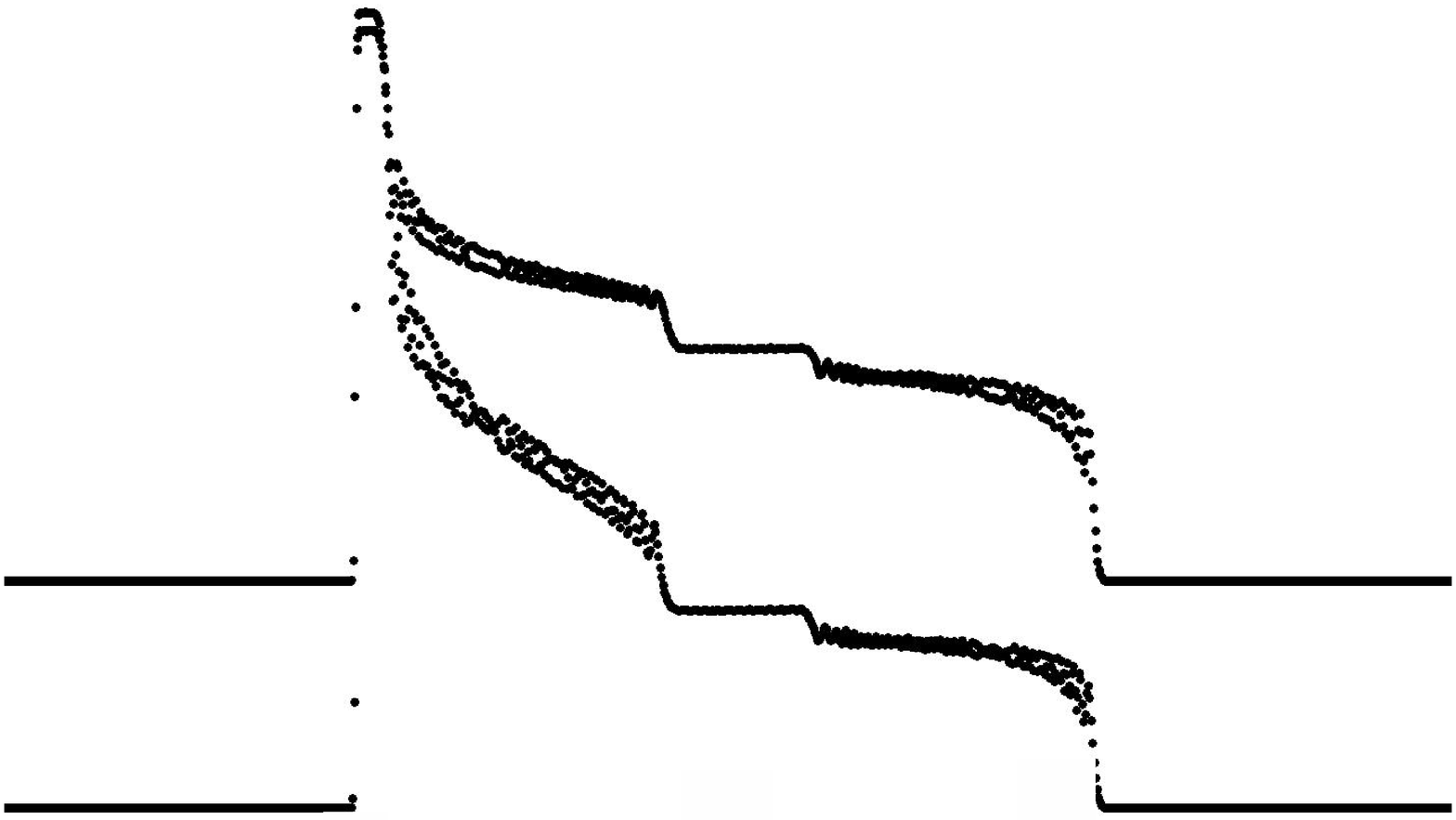}
        \caption{$t=500$}
    \end{subfigure}
\caption{Two snapshots at times $t=300$ (A) and $t=500$ (B) of the (numerically computed) 
         solution of a FPUT$-\alpha$ lattice for $\alpha=0.25$ with periodic background 
         initial condition.}
         \label{fpu_a_period}
\end{figure}
In  Figure \ref{fpu_a_period} we observe one soliton, three pure periodic regions and two modulated 
oscillation regions in between, very similar to the Toda case in Figure \ref{fig12}.
It should be added that we observe exactly the same behavior (qualitatively) in a plethora
of other situations. Folowing you can find a list of these cases
\begin{itemize}
\item
the Langmuir lattice
\item
the cube perturbed Langmuir chain at least for $\beta=0.01$ (or even smaller)
\item
$(2,1)-$Lennard-Jones potential for the values $(d,\varepsilon)=(10,10)$, $(1,10)$ and $(0.5,10)$
\item
Morse potential for the parameter values $(\gamma,\delta)=(4,1)$, $(4,0.5)$, $(4,0.25)$, $(4,0.01)$, 
$(8,1)$, $(8,0.5)$, $(8,0.25)$, $(8,0.1)$, $(8,0.1)$ and $(8,0.01)$
\item
both of the perturbed Toda potentials for \say{small} values ($0.01$ or smaller) of the parameters 
$\alpha$ and $\beta$ causing the perturbation
\end{itemize}
 
For a FPUT$-\beta$ chain with $\beta=0.01$ and periodic background, the simulations return the following figures
\begin{figure}[H]
    \centering
    \begin{subfigure}[b]{0.48\linewidth}        
        \centering
        \includegraphics[width=\linewidth]{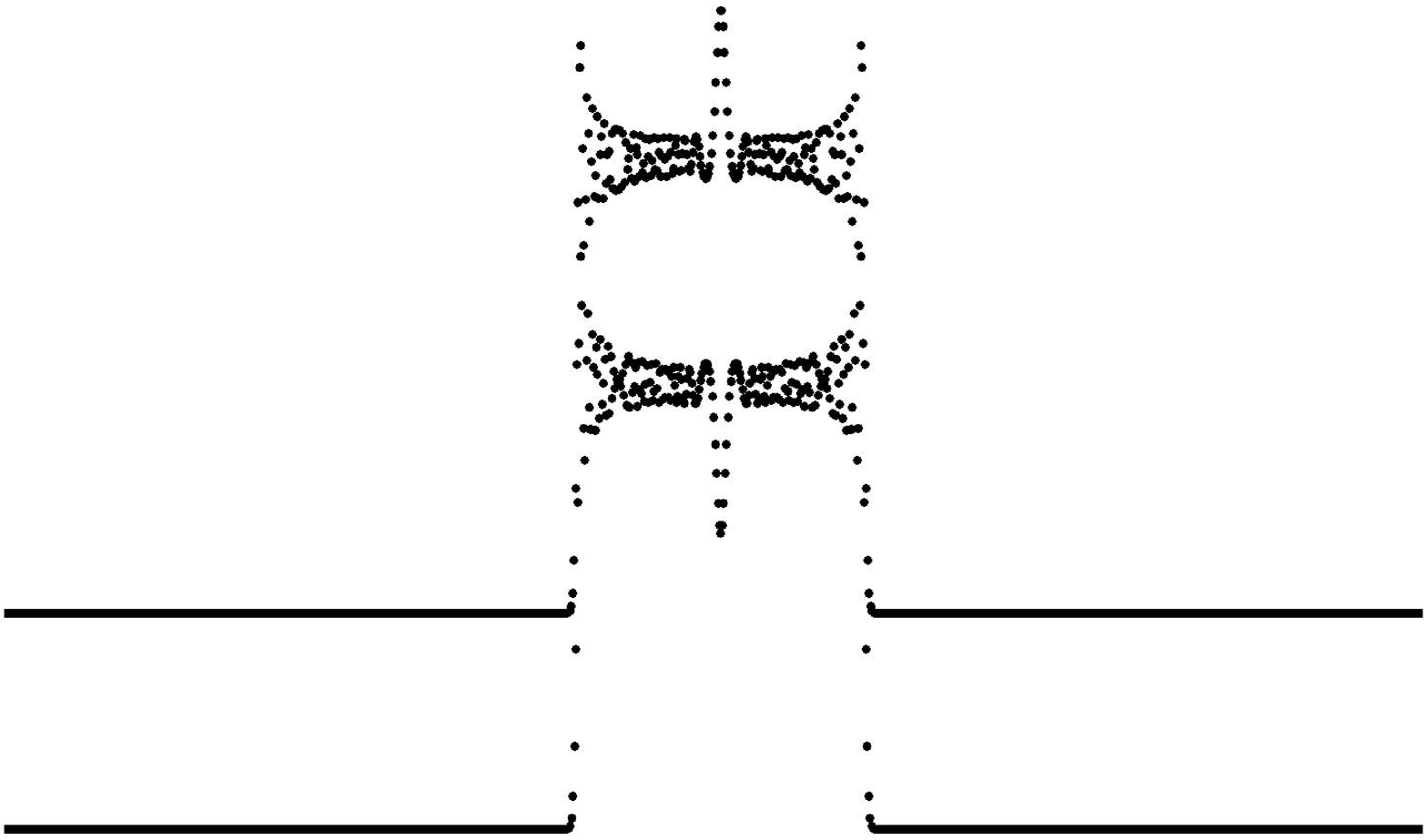}
        \caption{$t=200$}
    \end{subfigure}
    \begin{subfigure}[b]{0.48\linewidth}        
        \centering
        \includegraphics[width=\linewidth]{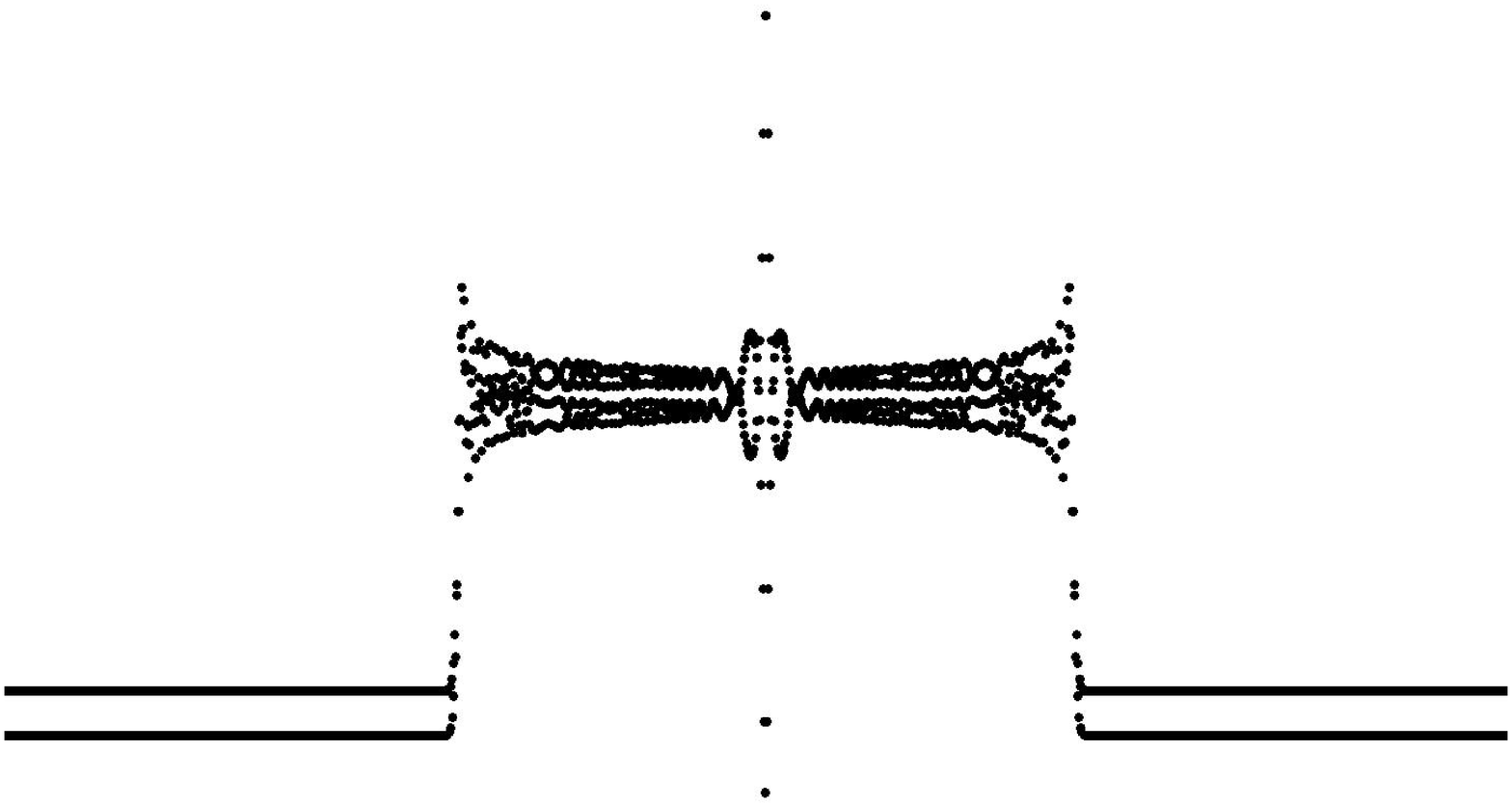}
        \caption{$t=400$}
    \end{subfigure}
\caption{Two snapshots at times $t=200$ (A) and $t=400$ (B) of the (numerically computed) 
         solution of a FPUT$-\beta$ lattice for $\beta=0.01$ with periodic background 
         initial condition.}
\end{figure}
In this case, there are two traveling solitons and one breather. There are also pure periodic regions 
in between. 

Although \say{small} perturbations of the completely integrable cases
still give  the same picture (i.e. pure periodicity plus modulations plus solitons), for larger perturbations 
this picture becomes more complicated. Even chaos can possibly appear.

\section{Conclusion}

We have investigated a soliton resolution conjecture for 
FPUT lattices in a constant or periodic background and we have presented  numerical computations supporting 
such a conjecture in the case of a FPUT lattice, for small perturbations of  completely
integrable one-dimensional lattices, but not necessarily for  larger perturbations. 
For the exact Toda, the computations have already been done many years ago in \cite{kt1} 
and complete proofs already exist (\cite{kt2}, \cite{krt}).

To make the conjecture  more precise:

\begin{src*}
Consider the solution of the initial value  problem for the FPUT nearest neighbour lattice in one dimension
which is a small perturbation of the linear harmonic lattice or the Toda lattice or in fact any integrable
lattice,  with initial data which
is asymptotically periodic in space. Then, we have the following facts  asymptotically:

1. The $(n,t)$-space,  splits into two kinds of regions separated by straight lines passing through
the origin.

2. There are regions of periodicity (the period being equal to the period of the  background),
and then there are regions where the PDE or lattice undergoes  modulated oscillations
with  large (order $1/t$) frequency and slowly varying (with $n/t$) amplitude and phase. Phenomena appear in two different scales
and are naturally expressed in two new variables: the \say{fast} one being $1/t$ and the \say{slow} one being $n/t$.
The regions of periodicity and the modulation regions are open cones bounded by half-lines (if we consider only positive times t)
emerging at the origin. There may also be solitons: travelling waves with constant shape and speed.
The soliton regions are small (in $1/t$)
regions around (some of) these half-lines. The slopes of the half-lines are the speeds of the solitons.

3. In the special case where the initial data background is constant the modulated oscillations region
does not occur.
\end{src*}

\begin{remark}
The conjecture is most certainly true when the forces between adjacent particles
render the lattices  integrable. Even though  proofs have not been produced for all possible
such lattices it is pretty clear that the inverse scattering -- Riemann-Hilbert methods will
produce the same results.  It is now also  confirmed numerically when a small extra term is added to these forces
even if integrability via inverse scattering is destroyed.
On the other hand general lattices away from  integrable cases above can exhibit a much less regular,
even chaotic behaviour.

\end{remark}

\begin{remark}
The above conclusion raises the following question.
How can  (and why) the soliton resolution conjecture be valid for any PDE of dispersive type and
not for all Hamiltonian lattices with forces between adjacent particles?

We admit that the  answer to this question eludes at this point!  
\end{remark}

\begin{remark}
We also believe that similar phenomena will appear in higher space dimensions.
But it remains to be seen what kind of coherent structures appear in place of the simple trivial background solitons. 
\end{remark}

\begin{remark}
Back in the last decade where the soliton resolution conjecture was first generalised to non-integrable NLS-type
equations it was only deemed realistic to consider a trivial background (\cite{t}).
In view of the recent flurry of activity involving \say{rogue wave} phenomena, which only exist for non-trivial backgrounds
and have only been rigorously treated in the case of a periodic background, we feel that a generalisation to
periodic background 
deserves to be considered. A background  with an indefinite reservoir of energy is very realistic when one considers, say, the
huge oceans.
 \end{remark}

\appendix

\section{Long Time Asymptotics of the Periodic Toda Lattice under Short-Range Perturbations and the Riemann-Hilbert method}
\label{asympt}

We summarise here the most important results of \cite{kt2}.
Consider the doubly infinite Toda lattice in Flaschka's variables
\be \label{TLpert}
\aligned
\dot b(n,t) &= 2\Big[a(n,t)^2 -a(n-1,t)^2\Big],\\
\dot a(n,t) &= a(n,t)\Big[b(n+1,t) -b(n,t)\Big],\quad (n,t) \in \Z \times \R
\endaligned
\ee
where the dot denotes differentiation with respect to time
and $a(n,t)$, $b(n,t)$
are the Flaschka
variables
\be \label{Flaschka}
\aligned
a(n,t) &= \frac{1}{2}\exp\Big\{\tfrac{1}{2}\big[x(n,t) -x(n+1,t)\big]\Big\}\\
b(n,t) &= -\frac{1}{2} \dot x(n,t),
\endaligned
\ee

In this appendix we will consider a periodic algebro-geometric background
solution $(a_q,b_q)$ to be described in a while in the next paragraph, plus
a short-range perturbation $(a,b)$ satisfying
\be \label{decay}
\sum_{n\in\Z}\Big[ n^6 \big(|a(n,t) - a_q(n,t)| + |b(n,t)- b_q(n,t)|\big)\Big] < \infty
\ee
for $t=0$  and hence for all   $t\in\R$. The
perturbed solution can be analysed  with the help  of the inverse scattering transform in a periodic background 
(\cite{emt}).

To fix our background solution, consider a hyperelliptic Riemann surface of genus $g$
with real moduli $E_0, E_1,..., E_{2g+1}$. Choose a Dirichlet divisor $\dimuz$ and introduce
\be
\ulz(n,t) = \hAmap(\infty_+) - \hamap(\dimuz) - n\ul{\hat A}_{\infty_-}(\infty_+)
+ t\ul{U}_0 - \hvrc \in \C^g,
\ee
where $\Amap$ ($\amap$) is Abel's map (for divisors) and $\vrc$, $\ul{U}_0$ are
some properly defined constants.
Then our background solution is given in terms
of Riemann theta functions  by
\begin{align} \nn
a_q(n,t)^2 &= \ti{a}^2 \frac{\theta(\ulz(n+1,t)) \theta(\ulz(n-1,t))}{\theta(
\ulz(n,t))^2},\\
b_q(n,t) &= \tilde{b} + \frac{1}{2}
\frac{d}{dt} \log\Bigg[\frac{\theta(\ulz(n,t)) }{\theta(\ulz(n-1,t))}\Bigg],
\end{align}
where $\ti{a}$, $\tilde{b} \in\R$ are again some constants.

We can of course view this hyperelliptic Riemann surface as formed by cutting and pasting
two copies of the complex plane along bands. Having this picture
in mind, we denote the standard projection
to the complex plane by $\boldsymbol{\pi}$.

Assume for simplicity that the Jacobi operator
\be
H(t) f(n) = a(n,t) f(n+1) + a(n-1,t) f(n-1) + b(n,t) f(n), \quad f\in\ell^2(\Z),
\ee
corresponding to the perturbed problem \eqref{TLpert} has no eigenvalues.
Then,  for long times the perturbed Toda lattice
is  asymptotically  close to the following limiting lattice
defined by
\be \label{limlat}
\aligned
\prod_{j=n}^{\infty} \Bigg[\frac{a_l(j,t)}{a_q(j,t)} \Bigg]^2 = 
\frac{\theta(\ulz(n,t))}{\theta(\ulz(n-1,t))} &
\frac{\theta(\ulz(n-1,t)+ \ul{\delta}(n,t))}{\theta(\ulz(n,t)+\ul{\delta}(n,t))} \times\\
& \times \exp\left( \frac{1}{2\pi\I} \int_{C(n/t)}
\log (1-|R|^2) \om_{\infty_+\, \infty_-}\right),\\
\delta_\ell(n,t)= \frac{1}{2\pi\I} &\int_{C(n/t)} \log (1-|R|^2) \zeta_\ell,
\endaligned
\ee
where $R$ is the  reflection coefficient defined when considering  scattering
with respect to the periodic background (see \cite{kt2} for the actual definition; it encapsulates the
short range perturbation),
$\zeta_\ell$ is a canonical basis
of holomorphic differentials, $\om_{\infty_+\, \infty_-}$ is an Abelian differential of
the third kind defined in \eqref{ominfpm}, and $C(n/t)$ is a contour on the Riemann
surface. More specific, $C(n/t)$ is obtained by taking the spectrum of the unperturbed
Jacobi operator $H_q$ between $-\infty$ and a special stationary phase point $z_j(n/t)$,
for the phase of the underlying Riemann--Hilbert problem (see below), and lifting it to the Riemann surface
(oriented such that the upper sheet lies to its left). The point $z_j(n/t)$ will move from
$-\infty$ to $+\infty$ as $n/t$ varies from $-\infty$ to $+\infty$.
 From the products above, one easily recovers $a_l(n,t)$.
More precisely, from \cite{kt2} we have the following:

\begin{theorem}\label{thmMain}
Let $C$ be any  (large) positive number and $\delta$ be any (small)
positive number.
Consider the  region
$D= \{(n,t): |\frac{n}{t}| < C \} $. Then one has
\be
\prod_{j=n}^{\infty} \frac{a_l(j,t)}{a(j,t)} \to 1
\ee
uniformly in $D$, as $t \to \infty$.
\end{theorem}

A similar theorem can be proved for the  velocities $b(n,t)$:

\begin{theorem}\label{thmMain2}
In the  region
$D= \{(n,t): |\frac{n}{t}| < C \},$
of Theorem ~\ref{thmMain} we also have
\be
\sum_{j=n}^\infty \big[ b_l(j,t) - b_q(j,t) \big] \to 0
\ee
uniformly in $D$, as $t \to \infty$, where $b_l$ is given by
\be
\aligned
\sum_{j=n}^\infty \big[ b_l(j,t) - b_q(j,t) \big] = &
\frac{1}{2\pi\I} \int_{C(n/t)} \log(1-|R|^2) \Omega_0\\
& {} + \frac{1}{2}\frac{d}{ds}
\log\left( \frac{\theta(\ulz(n,s) + \ul{\delta}(n,t) )}{\theta(\ulz(n,s))} \right) \Bigg|_{s=t}
\endaligned
\ee
and $\Omega_0$ is an Abelian differential of the second kind defined in \eqref{Om0}.
\end{theorem}

\begin{remark}

(i) It is easy to see how the asymptotic formulae above describe the picture
given by the numerics. Recall that the spectrum $\sig(H_q)$ of $H_q$ consists
of $g+1$ bands whose band edges are the branch points of the underlying
hyperelliptic Riemann surface. If $\frac{n}{t}$ is small enough, $z_j(n/t)$ is to the
left of all bands implying that $C(n/t)$ is empty and thus $\delta_\ell(n,t)=0$;
so we recover the purely periodic lattice.
At some value of  $\frac{n}{t}$  a stationary phase point first appears in the
first band of $\sig(H_q)$ and begins to move from
the left endpoint of the band towards the right endpoint of the band.
(More precisely we have a pair of stationary phase points $z_j$ and $z_j^*$, one in each sheet
of the hyperelliptic curve, with common projection $\boldsymbol{\pi}(z_j)$ on the complex plane.)
So $\delta_\ell(n,t)$ is now a non-zero quantity changing with $\frac{n}{t}$
and the asymptotic lattice has a slowly modulated non-zero phase.
Also the factor given by the exponential of the integral is
non-trivially changing with $\frac{n}{t}$ and contributes  to a
slowly modulated amplitude. Then, after the
stationary phase point leaves the first band
there is a range of $\frac{n}{t}$ for which
no stationary phase point appears in the spectrum $\sig(H_q)$, hence the phase shift
$\delta_\ell(n,t)$ and the integral remain constant, so the asymptotic lattice is periodic
(but with a non-zero phase shift). Eventually a stationary phase point
appears in the second band, so a new modulation appears and so on.
Finally, when $\frac{n}{t}$ is large enough, so that all bands have been
traversed by the stationary phase point(s), the asymptotic lattice is
again periodic. Periodicity properties of theta functions
easily show that phase shift  is actually cancelled by the exponential
of the integral and we recover the original periodic lattice with no
phase shift at all.

(ii) If eigenvalues are present one can apply appropriate Darboux transformations
to add the effect of such eigenvalues. Alternatively one can modify the
Riemann-Hilbert problem by adding small
circles around the extra poles coming from the eigenvalues
and applying some of the methods in \cite{dkkz}.
What we then see asymptotically is
travelling solitons in a periodic background. Note that this will change the
asymptotics on one side. More precisely  we have (see \cite{krt}) the following formulae:

\begin{theorem}\label{thm:asym}
Assume \eqref{decay} and denote the eigenvalues of the Jacobi operatror by $\rho_k, ~~~k=1,...., N$.
Let  $c_k= v(\rho_k)$ (the velocity of the $k^{th}$ soliton) defined via
\be
v(\lam) = \lim_{\eps\to 0} \frac{-\re \int_{E_0}^{(\lam+\I\eps,+)}\! \Omega_0}{ \re \int_{E_0}^{(\lam+\I\eps,+)}\! \omega_{\infty_+\,\infty_-}},
\ee
where $\Omega_0$ is an Abelian differential of the second kind defined in \eqref{Om0}
and $ \om_{\infty_+\, \infty_-} $ is
the Abelian differential of the third kind with poles at $\infty_+$ and $\infty_-$ defined in \eqref{ominfpm}.
Also let $\eps > 0$ sufficiently small such that the intervals $[c_k-\eps,c_k+\eps]$, $1\le k \le N$, are disjoint and lie
inside $v\big(\R\backslash\sig(H_q)\big)$.
Then the asymptotics in the soliton region, $\{ (n,t) |\, \zeta(n/t) \in\R\backslash\sig(H_q)\}$,
are as follows:
\begin{itemize}
\item
if $|\frac{n}{t} - c_k|<\eps$ for some $k$, the solution is asymptotically given by a one-soliton
solution on top of the limiting lattice:
\begin{align}\nn
\prod_{j=n}^{\infty} \frac{a(j,t)}{a_l(j,t)} &= \left(
\sqrt{\frac{c_{l,\gam_k(n,t)}(\rho_k,n-1,t)}{c_{l,\gam_k(n,t)}(\rho_k,n,t)}} + O(t^{-l}) \right),\\
\sum_{j=n+1}^\infty b(j,t) - b_l(j,t)&= -\gam_k(n,t) \frac{a_l(n,t) \psi_l(\rho_k,n,t) \psi_l(\rho_k,n+1,t)}{2 c_{l,\gam_k(n,t)}(\rho_k,n,t)}
+ O(t^{-l}),
\end{align}
for any $l \geq 1$, where
\begin{equation}
c_{l,\gam}(\rho,n,t) = 1 + \gam\!\! \sum_{j=n+1}^\infty \psi_{l,+}(\rho,j,t)^2
\end{equation}
and
\be\label{eq:gamshift}
\gam_k(n,t) = \gam_k \frac{T(\rho_k^*,n,t)}{T(\rho_k,n,t)}.
\ee
\item
if $|\frac{n}{t} -c_k| \geq \eps$, for all $k$, the solution is asymptotically close to the limiting lattice:
\begin{align}\nn
\prod_{j=n}^{\infty} \frac{a(j,t)}{a_l(j,t)} &= 1 + O(t^{-l}),\\
\sum_{j=n+1}^\infty b(j,t) - b_l(j,t) &= O(t^{-l}),
\end{align}
for any $l \geq 1$.
\end{itemize}
\end{theorem}

Here $\psi_l(p,n,t)$ is the Baker-Akhiezer function (cf.\ Section~\ref{secAG}) 
corresponding to the limiting lattice defined above.
The suffix $\pm$ refers to the restriction on the $\pm$ sheet and the star denotes sheet flipping.
$T$ is the transition coefficient
defined when considering  scattering
with respect to the periodic background.

(iii) It is very easy to also show that in any region
$|\frac{n}{t}|> C$, one has
\be
\prod_{j=n}^{\infty} \frac{a_l(j,t)}{a(j,t)} \to 1
\ee
uniformly in $t$, as $n \to \infty$.

\end{remark}

By dividing in \eqref{limlat} one recovers the $a(n,t)$.
It follows from the theorem above that
\be
|a(n,t) - a_l(n,t)| \to 0
\ee
uniformly in $D$, as $t \to \infty$.
In other words,
the perturbed  Toda lattice
is  asymptotically close to the  limiting  lattice above.

The proof  is based on a stationary phase type argument.
One reduces the given Riemann-Hilbert problem to a localised parametrix Riemann-Hilbert problem.
This is done via the solution of a scalar global Riemann-Hilbert problem
which is solved explicitly with the help  of the Riemann-Roch theorem.
The reduction to a
localised parametrix Riemann-Hilbert problem  is
done with the help of a theorem reducing general
Riemann-Hilbert problems to singular integral equations. (A generalized Cauchy transform
is defined appropriately for each Riemann surface.)
The localised parametrix Riemann-Hilbert problem  is  solved explicitly
in terms of parabolic cylinder functions. The argument follows \cite{dz} up to
a point but also extends the theory of Riemann-Hilbert problems for
Riemann surfaces. The right (well-posed) Riemann-Hilbert factorisation
problems are no more holomorphic but instead
have a number of poles equal to the genus of the surface.

\section{Algebro-geometric quasi-periodic finite-gap solutions}
\label{secAG}

We present  some facts on our background
solution $(a_q,b_q)$ which we want to choose from the class of
algebro-geometric quasi-periodic finite-gap solutions, that is the
class of stationary solutions of the Toda hierarchy.
In particular, this class contains all periodic solutions. We will
use the same notation as in \cite{tjac}, where we also refer to for proofs.

To set the stage let $\M$ be the Riemann surface associated with the following function
\begin{equation}
\Rg{z}, \qquad R_{2g+2}(z) = \prod_{j=0}^{2g+1} (z-E_j), \qquad
E_0 < E_1 < \cdots < E_{2g+1},
\end{equation}
$g\in \N$. $\M$ is a compact, hyperelliptic Riemann surface of genus $g$.
We will choose $\Rg{z}$ as the fixed branch
\begin{equation}
\Rg{z} = -\prod_{j=0}^{2g+1} \sqrt{z-E_j},
\end{equation}
where $\sqrt{.}$ is the standard root with branch cut along $(-\infty,0)$.

A point on $\M$ is denoted by
$p = \big(z, \pm \Rg{z}\big) = (z, \pm)$, $z \in \C$, or $p = (\infty,\pm) = \infty_\pm$, and
the projection onto $\C \cup \{\infty\}$ by $\boldsymbol{\pi}(p) = z$.
The points $\{(E_{j}, 0), 0 \leq j \leq 2 g+1\} \subseteq \M$ are
called branch points and the sets
\begin{equation}
\Pi_{\pm} = \{ \big(z, \pm \Rg{z}\big) \big| z \in \C\setminus
\bigcup_{j=0}^g[E_{2j}, E_{2j+1}]\} \subset \M
\end{equation}
are called upper, lower sheet, respectively.

Let $\{a_j, b_j\}_{j=1}^g$ be loops on the surface $\M$ representing the
canonical generators of the fundamental group $\boldsymbol{\pi}_1(\M)$. We require
$a_j$ to surround the points $E_{2j-1}$, $E_{2j}$ (thereby changing sheets
twice) and $b_j$ to surround $E_0$, $E_{2j-1}$ counterclockwise on the
upper sheet, with pairwise intersection indices given by
\begin{equation}
a_i \circ a_j= b_i \circ b_j = 0, \qquad a_i \circ b_j = \delta_{i,j},
\qquad 1 \leq i, j \leq g.
\end{equation}
The corresponding canonical basis $\{\zeta_j\}_{j=1}^g$ for the space of
holomorphic differentials can be constructed by
\begin{equation}
\underline{\zeta} = \sum_{j=1}^g \underline{c}(j)
\frac{\boldsymbol{\pi}^{j-1}d\boldsymbol{\pi}}{R_{2g+2}^{1/2}},
\end{equation}
where the constants $\underline{c}(.)$ are given by
\be\label{defcjk}
c_j(k) = C_{jk}^{-1}, \qquad
C_{jk} = \int_{a_k} \frac{\boldsymbol{\pi}^{j-1}d\boldsymbol{\pi}}{R_{2g+2}^{1/2}} =
2 \int_{E_{2k-1}}^{E_{2k}} \frac{z^{j-1}dz}{\Rg{z}} \in
\R.
\ee
The differentials fulfill
\begin{equation} \label{deftau}
\int_{a_j} \zeta_k = \delta_{j,k}, \qquad \int_{b_j} \zeta_k = \tau_{j,k},
\qquad \tau_{j,k} = \tau_{k, j}, \qquad 1 \leq j, k \leq g.
\end{equation}

Now pick $g$ numbers (the Dirichlet eigenvalues)
\be
(\hat{\mu}_j)_{j=1}^g = (\mu_j, \sigma_j)_{j=1}^g
\ee
whose projections lie in the spectral gaps, that is, $\mu_j\in[E_{2j-1},E_{2j}]$.
Associated with these numbers is the divisor $\dimuz$ which
is one at the points $\hat{\mu}_j$  and zero else. Using this divisor we
introduce
\begin{align} \nn
\ulz(p,n,t) &= \hAmap(p) - \hamap(\dimuz) - n\ul{\hat A}_{\infty_-}(\infty_+)
+ t\ul{U}_0 - \hvrc \in \C^g, \\
\ulz(n,t) &= \ulz(\infty_+,n,t),
\end{align}
where $\vrc$ is the vector of Riemann constants
\begin{equation}
\hat{\Xi}_{p_0,j} = \frac{j+ \sum_{k=1}^g \tau_{j,k}}{2},
\qquad p_0=(E_0,0),
\end{equation}
$\ul{U}_0$ are the \textit{b}-periods of the Abelian differential $\Omega_0$ defined below,
and $\Amap$ ($\amap$) is Abel's map (for divisors). The hat indicates that we
regard it as a (single-valued) map from $\hat{\M}$ (the fundamental polygon
associated with $\M$ by cutting along the $a$ and $b$ cycles) to $\C^g$.
We recall that the function $\theta(\ulz(p,n,t))$ has precisely $g$ zeros
$\hmu_j(n,t)$ (with $\hmu_j(0,0)=\hmu_j$), where $\theta(\ul{z})$ is the
Riemann theta function of $\M$.

Then our background solution is given by
\begin{align} \nn
a_q(n,t)^2 &= \ti{a}^2 \frac{\theta(\ulz(n+1,t)) \theta(\ulz(n-1,t))}{\theta(
\ulz(n,t))^2},\\ \label{imfab}
b_q(n,t) &= \tilde{b} + \frac{1}{2}
\frac{d}{dt} \log\Big[\frac{\theta(\ulz(n,t)) }{\theta(\ulz(n-1,t))}\Big].
\end{align}
The constants $\ti{a}$, $\tilde{b}$ depend only on the Riemann surface
(see \cite{tjac} section $9.2$).

Introduce the time dependent Baker-Akhiezer function
\begin{align}\label{defpsiq}
\psi_q(p,n,t) &= C(n,0,t) \frac{\theta (\ulz(p,n,t))}{\theta(\ulz (p,0,0))}
\exp \Big( n \int_{E_0}^p \om_{\infty_+\, \infty_-} + t\int_{E_0}^p \Omega_0
\Big),
\end{align}
where $C(n,0,t)$ is real-valued,
\begin{equation}
C(n,0,t)^2 = \frac{ \theta(\ulz(0,0)) \theta(\ulz(-1,0))}
{\theta (\ulz (n,t))\theta (\ulz (n-1,t))},
\end{equation}
and the sign has to be chosen in accordance with $a_q(n,t)$.
Here
\be
\theta(\ulz) = \sum_{\ul{m} \in \Z^g} \exp \Big\{2 \pi \I \left( \spr{\ul{m}}{\ulz} +
\frac{\spr{\ul{m}}{\ul{\tau} \, \ul{m}}}{2}\right)\Big\},\qquad \ulz \in \C^g,
\ee
is the Riemann theta function associated with $\M$,
\be\label{ominfpm}
\om_{\infty_+\, \infty_-}= \frac{\prod_{j=1}^g (\boldsymbol{\pi} -\lambda_j) }{R_{2g+2}^{1/2}}d\boldsymbol{\pi}
\ee
is the Abelian differential of the third kind with poles at $\infty_+$ and $\infty_-$ and
\be\label{Om0}
\Omega_0 = \frac{\prod_{j=0}^g (\boldsymbol{\pi} - \ti\lambda_j) }{R_{2g+2}^{1/2}}d\boldsymbol{\pi},
\qquad \sum_{j=0}^g \ti\lambda_j = \frac{1}{2} \sum_{j=0}^{2g+1} E_j,
\ee
is the Abelian differential of the second kind with second order poles at
$\infty_+$ respectively $\infty_-$ (see \cite[Sects.~13.1, 13.2]{tjac}).
All Abelian differentials are normalized to have vanishing $a_j$ periods.

The Baker-Akhiezer function is a meromorphic function on $\M\setminus\{\infty_\pm\}$
with an essential singularity at $\infty_\pm$. The two branches are denoted by
\begin{equation}
\psi_{q,\pm}(z,n,t) = \psi_q(p,n,t), \qquad p=(z,\pm)
\end{equation}
and it satisfies
\begin{align}\nn
H_q(t) \psi_q(p,n,t) &= \boldsymbol{\pi}(p) \psi_q(p,n,t),\\
\frac{d}{dt} \psi_q(p,n,t) &= P_{q,2}(t) \psi_q(p,n,t),
\end{align}
where
\begin{align}
H_q(t) f(n) &= a_q(n,t) f(n+1) + a_q(n-1,t) f(n-1) + b_q(n,t) f(n),\\
P_{q,2}(t) f(n) &= a_q(n,t) f(n+1) - a_q(n-1,t) f(n-1)
\end{align}
are the operators from the Lax pair for the Toda lattice.

It is well known that the spectrum of $H_q(t)$ is time independent and
consists of $g+1$ bands
\begin{equation}
\sig(H_q) = \bigcup_{j=0}^g [E_{2j},E_{2j+1}].
\end{equation}

\section{MATLAB\textsuperscript{\textregistered} code}
\label{code}

Here we present the code used for our simulations. The main program we ran in 
MATLAB\textsuperscript{\textregistered} is\\
\texttt{
clear all; close all; clc\\
\%number of particles (a power of 2)\\
N=2048;\\
\%size of time-step\\
DT=1;\\
\%number of time-steps\\
TMAX=800;\\
\%discretization of time-inteval\\                 
tspan=$[$0:DT:TMAX$]$;\\    
\%test different tolerances, changing Reltol\\
options=odeset('Reltol',1e-4,'OutputFcn','odeplot','OutputSel',[1,2,N]);\\ 
\%define initial-condition vector\\
\%first N entries denote position \& last N entries velocity\\
b=zeros(2*N,1);\\
\%our two initial conditions\\
\%we uncomment only one of them each time we run this code\\
for I=1:N\\ 
\%zero background initial conditions\\
\%b(I)=exp(-((I-N/4)/4)\^{}2); b(I+N)=0;\\   
\%periodic background initial condition\\
\%b(I)=0; b(I+N)=(-1)\^{}I+2*(I==N/2);\\  
end\\
\%time integration method\\
$[$t,y$]$=ode45('diffsystem',tspan,b,options,N);\\
}

where  the function \texttt{diffsystem} is defined as follows\\
\texttt{
function $[$db$]$=diffsystem(t,b)\\
\%number of particles (a power of 2)\\
N=2048;\\
for K=1:N\\
D(K)=b(N+K);\\    
end\\
\%the function p in what follows represents the potential function we consider\\ 
\%in each case (e.g. Toda potential) and is defined in another file\\
D(N+1)=p(b(2)-b(1))-p(b(1)-b(N));\\
for L=2:N-1\\
D(N+L)=p(b(L+1)-b(L))-p(b(L)-b(L-1));\\
end\\
D(2*N)=p(b(1)-b(N))-p(b(N)-b(N-1));\\
db=D';\\
end
}

\end{document}